\documentclass[12pt,preprint]{aastex}
\usepackage{graphicx,subfigure,amsmath,natbib}
\usepackage[percent]{overpic}
\newcommand\gsim{\,\lower3pt\hbox{$\sim$}\llap{\raise2pt\hbox{$>$}}\,}
\newcommand\lsim{\,\lower3pt\hbox{$\sim$}\llap{\raise2pt\hbox{$<$}}\,}

\begin{document}
\title{A simulation of convective dynamo in the solar convective envelope:
maintenance of the solar-like differential rotation and emerging flux}

\author{Yuhong Fan and Fang Fang}
\affil{High Altitude Observatory, National Center for Atmospheric
Research, 3080 Center Green Drive, Boulder, CO 80301}

\begin{abstract}
We report the results of a magneto-hydrodynamic (MHD) simulation of a convective
dynamo in a model solar convective envelope driven by the solar radiative
diffusive heat flux.  The convective dynamo produces a large-scale mean
magnetic field that exhibits irregular cyclic behavior with oscillation time
scales ranging from about 5 to 15 years and undergoes irregular polarity
reversals.  The mean axisymmetric toroidal magnetic field is of
opposite signs in the two hemispheres and is concentrated at the bottom
of the convection zone.  The presence of the magnetic fields is found to
play an important role in the self-consistent maintenance of a solar-like
differential rotation in the convective dynamo model.  Without the magnetic
fields, the convective flows drive a differential rotation with a faster
rotating polar region. In the midst of magneto-convection, we found emergence
of strong super-equipartition flux bundles at the surface, exhibiting
properties that are similar to emerging solar active regions.
\end{abstract}

\keywords{magnetohydrodynamics(MHD) - Sun: dynamo - Sun: interior}

\section{Introduction}
Despite the turbulent nature of solar convection, the sun's large scale
magnetic field exhibits remarkable order and organization such as the
11-year sunspot cycle \citep[e.g.][]{Maunder:1922} and the Hale's polarity rule
of the bipolar active regions \citep{Hale:etal:1919,Hale:Nicholson:1925}.
In recent years, global fully dynamic three-dimensional (3D) convective dynamo
simulations have been making headway in producing the solar-like cyclic behavior
of the large scale magnetic field \citep[e.g.][]{Ghizaru:etal:2010,
Racine:etal:2011,Kapyla:etal:2012,Augustson:etal:2013}
and the self-consistent formation of buoyant, active region like emerging
tubes from dynamo generated strong toroidal fields
\citep{Nelson:etal:2011,Nelson:etal:2013,Nelson:etal:2014a}.
Most of these simulations have differential rotation with cylindrical
iso-rotation contours throughout the convection zone
\citep[see review by][]{Charbonneau:2013}, and some 
are considering rotation rate 3 times the solar rate.
In this paper we present a convective dynamo simulation
driven by the solar radiative diffusive
heat flux and maintains a differential rotation profile that resembles
more closely to the solar differential rotation in the convection zone
in terms of the pole-equator contrast and the more
conical iso-contours of rotation in the mid-latitude region.
The convective dynamo produces a large-scale mean magnetic field  
with irregular cyclic behavior and polarity reversals very similar
to a convective dynamo presented in \citet{Miesch:etal:2011}.
We demonstrate in this paper the important role the magnetic
fields play in the maintenance of the solar-like differential rotation.
We also show the emergence of strong super-equipartition flux tubes near
the surface that exhibit some properties similar to emerging solar active
regions.

\section{The Numerical Model\label{sec:model}}
We solve the following anelastic MHD equations
using a finite-difference spherical anelastic MHD code
\citep{Fan:2008,Fan:etal:2013}:
\begin{equation}
\nabla \cdot ( \rho_0 {\bf v} ) = 0,
\label{eq:continuity}
\end{equation}
\begin{equation}
\rho_0 \left [ \frac{\partial {\bf v}}{\partial t} + ({\bf v} \cdot \nabla )
{\bf v} \right ]
= 2 \rho_0 {\bf v} \times {\bf \Omega} - \nabla p_1 + \rho_1 {\bf g}
+ \frac{1}{4 \pi} ( \nabla \times {\bf B}) \times {\bf B}
+ \nabla \cdot {\cal D}
\label{eq:momentum}
\end{equation}
\begin{equation}
\rho_{0} T_0 \left [ \frac{\partial s_1}{\partial t}
 + ( {\bf v} \cdot \nabla ) (s_0 + s_1 ) \right ]
= \nabla \cdot ( K \rho_0 T_0 \nabla s_1 ) - ( {\cal D} \cdot \nabla )
\cdot {\bf v} + \frac{1}{4 \pi} \eta ( \nabla \times {\bf B} )^2
- \nabla \cdot {\bf F}_{\rm rad}
\label{eq:entropy1}
\end{equation}
\begin{equation}
\nabla \cdot {\bf B} = 0
\label{eq:divB}
\end{equation}
\begin{equation}
\frac{\partial {\bf B}}{\partial t} = \nabla \times ( {\bf v} \times {\bf B} )
- \nabla \times ( \eta \nabla \times {\bf B} ) ,
\label{eq:induction1}
\end{equation}
\begin{equation}
\frac{\rho_1}{\rho_0} = \frac{p_1}{p_0} - \frac{T_1}{T_0},
\label{eq:eqstate}
\end{equation}
\begin{equation}
\frac{s_1}{c_p} = \frac{T_1}{T_0} - \frac{\gamma -1}{\gamma}
\frac{p_1}{p_0} .
\label{eq:2ndthermlaw}
\end{equation}
In the above, $s_0 (r)$, $p_0 (r)$, $\rho_ 0 (r)$, $T_0 (r)$, and
${\bf g} = - g_0 (r) {\hat {\bf r}}$ denote
the profiles of entropy, pressure, density, temperature, and the gravitational
acceleration of a time-independent, reference state of hydrostatic equilibrium
and nearly adiabatic stratification, $c_p$ is the specific heat capacity at
constant pressure, $\gamma$ is the ratio of specific
heats, and ${\bf v}$, ${\bf B}$, $s_1$, $p_1$,
$\rho_1$, and $T_1$ are the velocity, magnetic field,
entropy, pressure, density, and temperature to be solved that describe the
changes from the reference state. ${\bf \Omega}$ denotes the solid body
rotation rate of the Sun and is the rotation rate of the frame of reference,
where $\Omega = 2.7 \times 10^{-6} {\rm rad} \; {\rm s}^{-1}$.
${\cal D}$ is the viscous stress tensor:
${\cal D}_{ij} = \rho_0 \nu \left [ S_{ij} - (2/3) ( \nabla \cdot
{\bf v} ) \delta_{ij} \right ]$,
where $\nu$ is the kinematic viscosity, $\delta_{ij}$ is the unit tensor, and
$S_{ij}$ is the strain rate tensor given by the following in spherical
polar coordinates:
\begin{equation}
S_{rr} = 2 \frac{\partial v_r}{\partial r}
\end{equation}
\begin{equation}
S_{\theta \theta} = \frac{2}{r} \frac{\partial v_{\theta}}{\partial \theta}
+ \frac{2 v_r}{r}
\end{equation}
\begin{equation}
S_{\phi \phi} = \frac{2}{r \sin \theta} \frac{\partial v_{\phi}}{\partial \phi}
+ \frac{2 v_r}{r} + \frac{2 v_{\theta}}{r \sin \theta } \cos \theta
\end{equation}
\begin{equation}
S_{r \theta} = S_{\theta r} = \frac{1}{r} \frac{\partial v_r}{\partial \theta}
+ r \frac{\partial}{\partial r} \left ( \frac{v_{\theta}}{r} \right )
\end{equation}
\begin{equation}
S_{\theta \phi } = S_{\phi \theta} = \frac{1}{r \sin \theta}
\frac{\partial v_{\theta}} {\partial \phi } + \frac{\sin \theta} {r}
\frac{\partial}{\partial \theta} \left ( \frac{v_{\phi}}{\sin \theta} \right )
\end{equation}
\begin{equation}
S_{\phi r} = S_{r \phi } = \frac{1}{r \sin \theta}
\frac{\partial v_r}{\partial \phi} + r \frac{\partial}{\partial r}
\left ( \frac{v_{\phi}}{r} \right ) .
\end{equation}
$K$ denotes the thermal diffusivity, and $\eta$ is the magnetic diffusivity.
In equation (\ref{eq:entropy1}),
\begin{equation}
{\bf F_{\rm rad}} = - \frac{ 16 \sigma_s {T_0}^3}{3 \kappa
\rho_0 } \frac{d T_0}{dr} \, {\hat {\bf r}}
\label{eq:frad}
\end{equation}
is the radiative diffusive heat flux,
where $\sigma_s$ is the Stephan-Boltzman constatn,
$\kappa$ is the Rosseland mean opacity.

The simulation domain is a partial spherical shell
with $r \in [r_i, r_o]$, spanning from $r_i=0.722 R_s$ at
the base of the convection zone (CZ) to $r_o = 0.971 R_s$
at about 20 Mm below the photosphere, where $R_s$ is the solar radius, 
$\theta \in [ \pi/2 - \Delta \theta, \pi/2 + \Delta \theta]$ with
$\Delta \theta = \pi / 3$, and $\phi \in [0, 2 \pi]$.
The domain is resolved by a grid with 96 grid points in $r$,
512 grid points in $\theta$, and 768  grid points in $\phi$.
J. Christensen-Dalsgaard's (JCD) solar model \citep{JCD:etal:1996}
is used for the reference profiles of $T_0$, $\rho_0$, $p_0$, $g_0$ in the
simulation domain. We assumed that $s_0 =0$ for the reference state.
The heating (the last term in eq. [\ref{eq:entropy1}])
due to the solar radiative diffusive heat flux drives a radial gradient
of $s_1$ that drives the convection.  We set the thermal diffusivity
$K = 3 \times 10^{13} \, {\rm cm}^2 \, {\rm s}^{-1}$, the viscosity $\nu =
10^{12} \, {\rm cm}^2 \, {\rm s}^{-1}$, and the magnetic diffusivity
$\eta = 10^{12} \, {\rm cm}^2 \, {\rm s}^{-1}$ at the top of the domain, and
they all decrease with depth following a $1/ \sqrt{\rho_0}$ profile.
The stratification of the domain includes approximately 4 density scale
heights between the top and the bottom, and thus the above diffusivities
decrease to $K = 4.02 \times 10^{12} \, {\rm cm}^2 \, {\rm s}^{-1}$, $\nu
= 1.34 \times 10^{11} \, {\rm cm}^2 \, {\rm s}^{-1}$, $\eta = 1.34 \times
10^{11} \, {\rm cm}^2 \, {\rm s}^{-1}$ at the bottom of the CZ domain.
The rationale for our choice of such depth dependent diffusivities
for the numerical experiments here is that, if the dominant spatial
scales of convection decreases with height, it may be expected
that more heat is transported by the unresolved scales as one moves towards
the top of the simulation domain, and hence the greater diffusivities there.
Furthermore, with a low magnetic diffusivity and viscosity in the deep CZ,
should buoyant magnetic structures develop, they would be able to better
preserve their magnetic buoyancy and rise.
Given the above diffusivities, the various diffusive time scales for the
simulation are estimated as follows. The viscous and magnetic diffusive time scales
$(\Delta r )^2 / \nu$ and $ (\Delta r)^2/ \eta$ range from about 71 years near
the bottom of CZ to about 10 years near the top, and the thermal diffusive time
scale $(\Delta r)^2 /K$ ranges from about 2.4 years near the bottom to about
0.3 years near the top, where we have used the depth of the CZ domain
$\Delta r$ for the estimate. 

We impose $\partial s_1 / \partial r = 0$ at the bottom and $s_1 = 0$ at the
top boundary.
We also impose a latitudinal gradient of entropy at the lower boundary:
\begin{equation}
\left ( \frac{\partial s_1 }{\partial \theta} \right )_{r_i} =
\frac{ \pi \Delta s_i }{ \Delta \theta } \sin \left (
\frac{ \pi (\theta - \pi / 2)}{ \Delta \theta } \right )
\label{eq:dsdth}
\end{equation}
where $\Delta s_i = 431.4 \: {\rm erg} \:
{\rm g}^{-1} \: {\rm K}^{-1}$, corresponding to a pole to equator
temperature difference of about $6.8$ K,
to represent the tachocline induced entropy variation that can break the
Taylor-Proudman constraint in the CZ \citep{Rempel:2005}.
At the two $\theta$ boundaries, $s_1$ is assumed symmetric.
The velocity boundary condition is non-penetrating and stress free at the
top, bottom and the two $\theta$-boundaries.
For the magnetic field we assume perfect conducting walls for the bottom and
the $\theta$-boundaries and radial field at the top boundary.
All quantities are naturally periodic at the $\phi$ boundaries.

For the initial state, we specify the initial $s_1$ such that its
horizontal average: $<\!\! s_1 \!\!>_{t=0}$, satisfies:
\begin{equation}
K \rho_0 T_0 \frac{d <\!\! s_1 \!\! >_{t=0}}{d r} = \frac{L_s}{4 \pi r^2} - F_{\rm rad},
\label{eq:s1init}
\end{equation}
where $L_s$ is the solar luminosity, $F_{\rm rad}$ is the absolute magnitude of
${\bf F}_{\rm rad}$ given in equation (\ref{eq:frad}), and at the lower
boundary $r_i$, $L_s / 4 \pi r^2 = F_{\rm rad}$.
Equation (\ref{eq:s1init}) lets the initial thermal conduction together with
radiative diffusion completely carry the solar luminosity, which sets up
an initial unstable entropy gradient $<\!\! s_1 \!\! >_{t=0}$.
We start the simulation with a small initial seed magnetic and velocity field
and let the magneto-convection evolve to a statistical steady state.

\section{Results \label{sec:results}}
\subsection{Overview of the convective dynamo \label{subsec:overview}}
\begin{figure}
\centering
\includegraphics[width=0.8\textwidth]{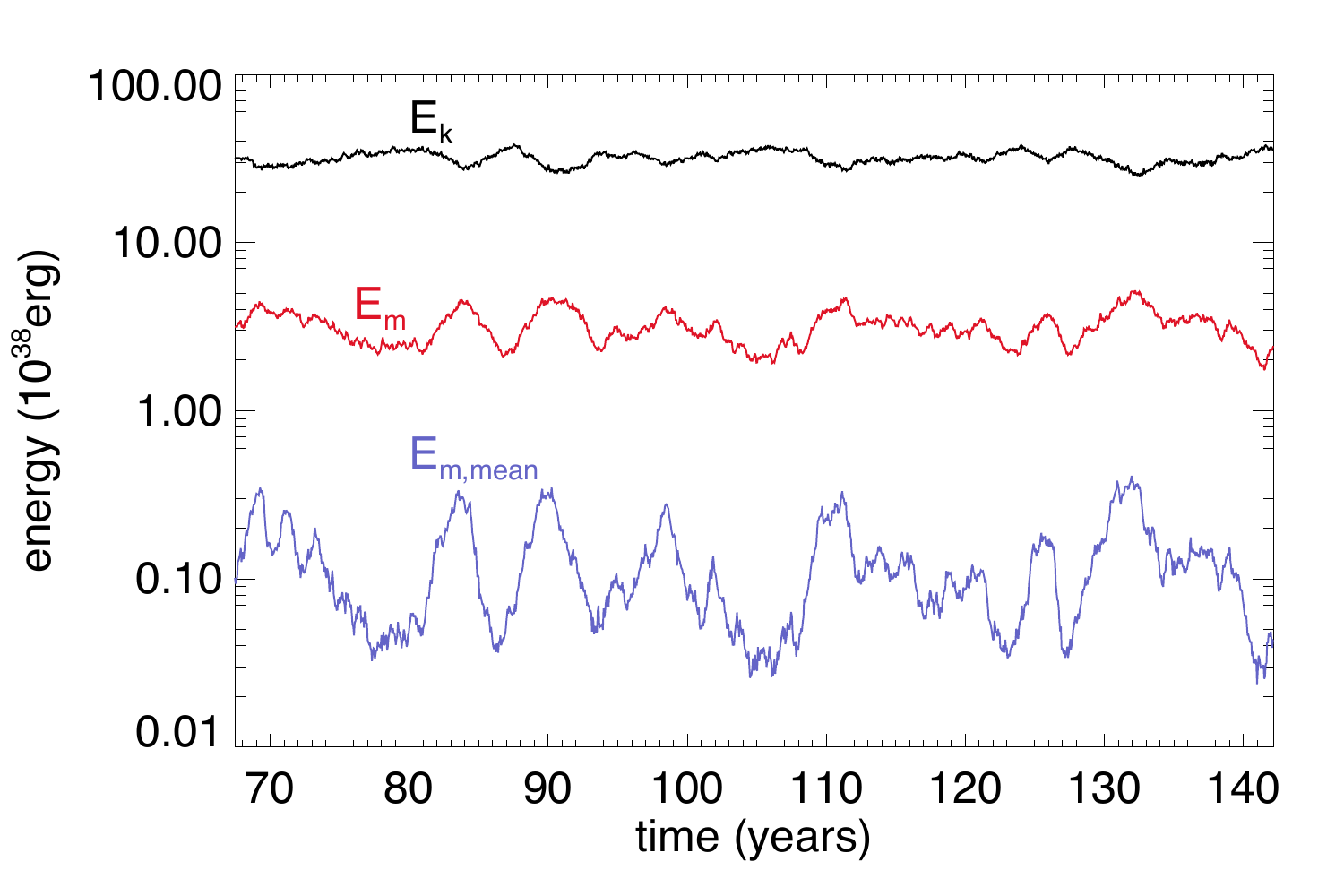}
\caption{The temporal variation of the total kinetic energy ($E_k$),
total magnetic energy ($E_m$), and the azimuthally averaged mean
magnetic field energy ($E_{\rm m, mean}$) of the statistically steady
convective flows in the simulation domain.}
\label{fig_energy}
\end{figure}

Figure \ref{fig_energy} shows the magnetic and kinetic
energies in the statistically steady convective flows in the simulation
domain over a time span of about 74 years.
The total magnetic energy $E_m$ maintained by the dynamo
is about 10\% of the total kinetic energy $E_k$ of the convective envelope.
The energy of the azimuthally averaged (mean) magnetic field $E_{\rm m, mean}$
only constitutes a small fraction of $E_m$,
oscillating from about 1\% to 10\% of $E_m$.

\begin{figure}
\centering
\includegraphics[width=0.8\textwidth]{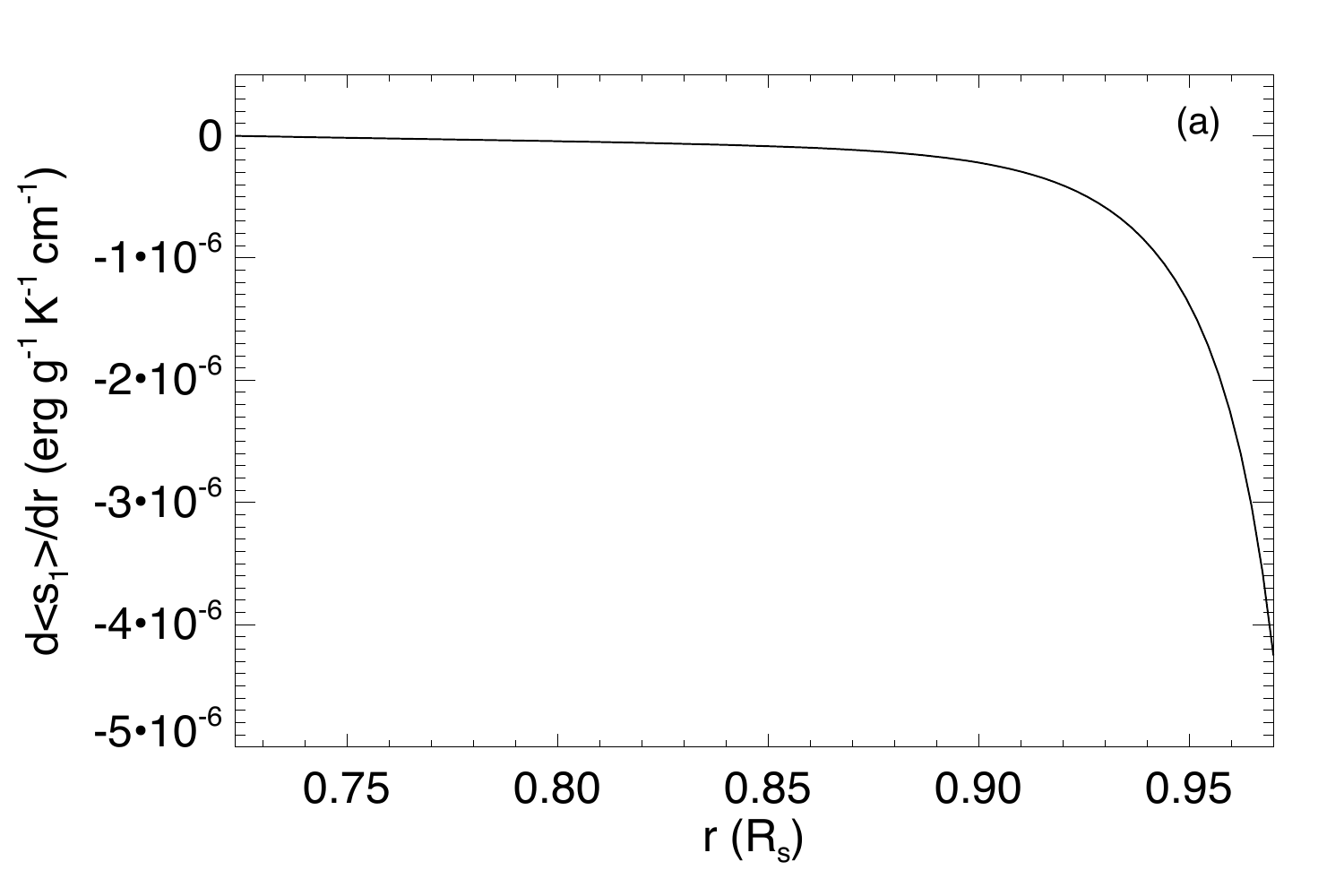}
\caption{The depth variation of the mean entropy gradient
established in the statistical steady state of the dynamo
 simulation}
\label{fig_dsdr}
\end{figure}

Figure \ref{fig_dsdr} shows the depth variation of the mean entropy
gradient established in the CZ domain in the statistical steady state.
The entropy gradient reaches a value of about $4.26 \times 10^{-6} \,
{\rm erg^{-1} K^{-1} cm^{-1}}$ at the top boundary at about $0.97 R_s$,
which is of a similar order of magnitude as the entropy gradient 
($\sim 10^{-5} \, {\rm erg^{-1} K^{-1} cm^{-1}}$) at this depth in the solar
model of JCD. Figure \ref{fig_heatfluxes} shows the various horizontally
integrated energy fluxes (normalized to the solar luminosity $L_s$)
through the domain as a function of radius established in the statistical
steady state.  These are respectively, the integrated radiative diffusive
heat flux (red curve): $L_{\rm rad} = 4 \pi r^2 F_{\rm rad}$,
the convective enthalpy flux (black curve):
\begin{equation}
L_{\rm conv} = 4 \pi r^2 \rho_0 c_p < \! \! v_r T_1 \! \!>
\label{eq:Lconv}
\end{equation}
the conductive energy flux by thermal diffusion (yellow curve):
\begin{equation}
L_{\rm cond} = - 4 \pi r^2
K \rho_0 T_0 \frac{d <\!\! s_1 \! \!>}{d r},
\label{eq:Lcond}
\end{equation}
the kinetic energy flux (blue curve):
\begin{equation}
L_{\rm kin} = 4 \pi r^2 \frac{\rho_0}{2} <\! \! v^2 v_r \!\!>
\label{eq:Lkin}
\end{equation}
the viscous energy flux (black dashed curve):
\begin{equation}
L_{\rm vis} = - 4 \pi r^2 < \! \! v_i {\cal D}_{ir} \! \! >
\label{eq:Lvis}
\end{equation}
the Poynting flux (green curve):
\begin{equation}
L_{\rm poyn}= - 4 \pi r^2 < \! \! \left ( \frac{1}{4 \pi} ({\bf v} \times {\bf B}) \times {\bf B} \right )_r \! \!> ,
\label{eq:Lpoyn}
\end{equation}
the resistive energy flux (cyan curve):
\begin{equation}
L_{\rm res}= 4 \pi r^2 < \! \! \left ( \frac{1}{4 \pi} ( \eta \nabla  \times {\bf B}) \times {\bf B} \right )_r \! \!> ,
\label{eq:Lresis}
\end{equation}
and the sum of all the energy fluxes, $L_{\rm tot}$, is shown as the dash-dotted curve in
Figure \ref{fig_heatfluxes}.
In the above $< >$ denotes averaging over the spherical shell surface and time.
The gradual decline of $L_{\rm tot}$ reflects a numerical deviation from exact energy
conservation, which is mainly caused by the numerical diffusion of the magnetic field
due to the Alfv\'en wave-upwind scheme used for advancing the induction equation
\citep{Fan:2008, Stone:Norman:1992:b}. This numerical dissipation of magnetic energy
is not being put back into the thermal energy in the entropy equation and
results in a loss of the total energy, and hence a loss of about $ 13$\% of the total
energy flux exiting the domain at the top compared to the total energy flux ($L_s$) entering
the domain from the bottom. We note that the explicit resistive dissipation of the magnetic
field (due to $\eta$), and both the explicit viscous dissipation (due to $\nu$) and the
numerical diffusion of momentum are put into the thermal energy in the entropy equation
as resistive and viscous heating to maintain energy conservation.  From
Figure \ref{fig_heatfluxes}, it can be seen that the enthalpy flux of the resolved convection
transports about 66\% of the solar luminosity in the middle of the CZ, and due to the high
thermal diffusivity $K$, thermal conduction also transports a substantial fraction
of the solar luminosity (about 36\% at the middle of the CZ).
The kinetic energy flux of the convective flows is downward and peaks at about 16\%
of the solar luminosity. The energy fluxes due to the Poynting flux (mostly downward),
resistive, and viscous transport are all much smaller.

\begin{figure}
\centering
\includegraphics[width=0.8\textwidth]{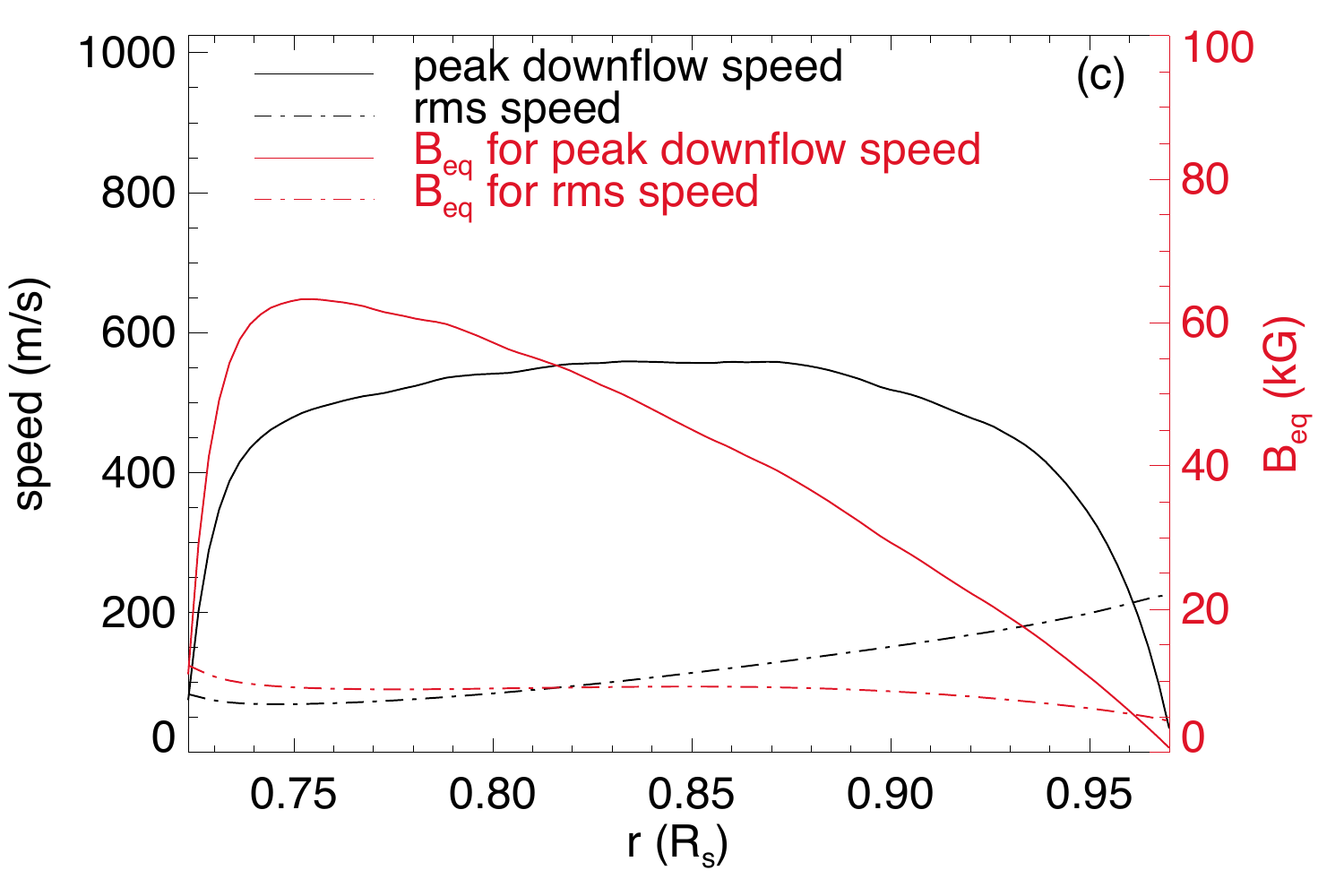}
\caption{The peak downflow speed, the RMS convective speed, and their
corresponding equipartition magnetic field strenghs as a function of depth.}
\label{fig_vbeq}
\end{figure}

Figure \ref{fig_vbeq} shows the depth variation of the peak
downflow (solid black curve), and the r.m.s. speed $v_{\rm rms}$
(dash-dotted black curve), of the statistical steady convective flows
in the domain.  
Note in computing $v_{\rm rms}$, we take out the azimuthally averaged velocity
components and only sum up the azimuthally fluctuating parts of the velocity
components.
Also shown are the corresponding magnetic field strength in equipartition with
the peak downflow speed (solid red curve) and the r.m.s. speed
(dash-dotted red curve).
It can be seen that the equipartition field strength $B_{\rm eq}$
corresponding to the peak down flow speed reaches $\approx 63$ kG, while
$B_{\rm eq}$ corresponding to the r.m.s. speed is $\sim 10$ kG for the deep
and mid convection zone, and decrease to about $5000$ G near the top
boundary at about $r = 0.971 R_s$.  Following \citet{Kapyla:etal:2012},
we compute the following non-dimensional numbers characterizing the
convective flows.
The Reynolds number
$ R_e = u_{\rm rms} / \nu k_f $ ranges from about 130 at the bottom
to about 50 at the top, and with a mid convection zone value of about 128,
where $k_f = 2 \pi / (r_o - r_i )$ and $u_{\rm rms} = ((3/2)< \! {v_r}^2 +
{v_{\theta}}^2 \! > )$ is the r.m.s. velocity averaged over each depth,
omitting the
contribution from the azimuthal velocity. The Coriolis number  
$C_O = ( 2 \Omega / u_{\rm rms, all} \, k_f ) = 1.3$,
where $u_{\rm rms, all} =
((3/2)< \! {v_r}^2 + {v_{\theta}}^2 \! > )$ with the averaging $<>$ done
for the entire domain. We can compare the values of these non-dimensional
numbers with the corresponding ones
in \citet{Kapyla:etal:2012}: $R_e = 36$
and $C_O = 7.6$. It appears the convective flow in our dynamo simulation
is moderately more turbulent as characterized by the larger $R_e$, especially
in the deeper layers of the CZ. Our Coriolis number $C_O$ is significantly
lower, indicating that our convective dynamo is operating in a significantly
less rotationally dominant regime.  If we were to scale their typical
r.m.s. velocity to be similar to ours $u_{\rm rms, all} \approx 100 $ m/s,
then their $C_O$ would imply a significantly more rapidly rotating stellar
envelope (with the solar CZ depth) at about 5 times the solar rotation rate.

\begin{figure}
\centering
\includegraphics[width=0.95\textwidth]{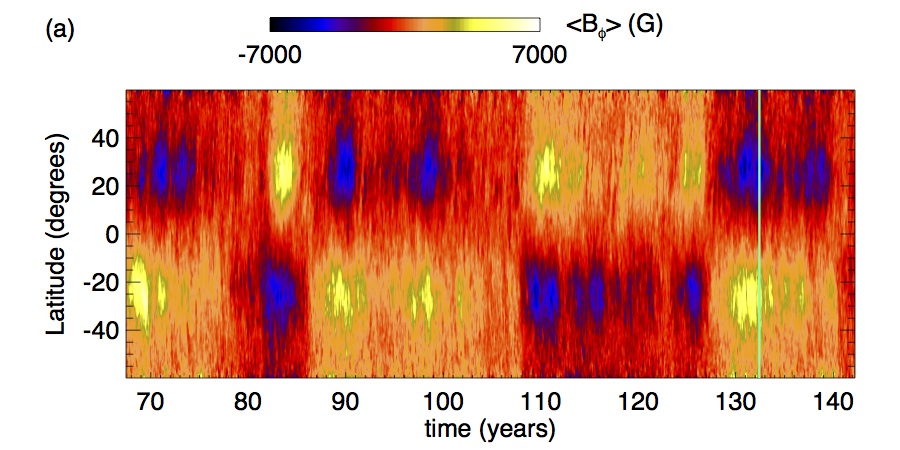} \\
\includegraphics[width=0.3\textwidth]{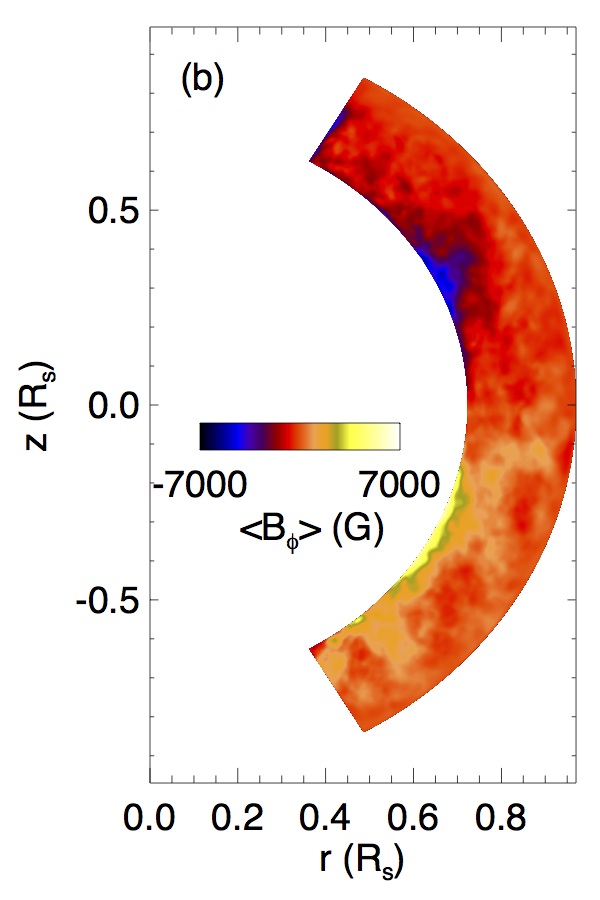}
\includegraphics[width=0.65\textwidth]{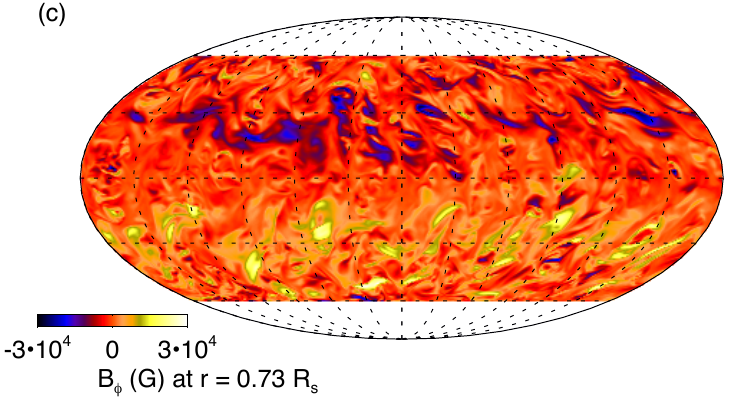}
\caption{(a) The latitude-time variation of the mean (azimuthally averaged)
toroidal magnetic field at a depth ($r=0.73 R_s$) near the bottom of the
CZ. (b) Azimuthally averaged toroidal magnetic field
distribution in the meridional plane at the time marked by the green line
in panel (a). (c) A shell slice of the toroidal magnetic field at a
depth ($r=0.73 R_s$) near the bottom of the CZ at the same
time marked by the green line in panel (a).}
\label{fig_cycles}
\end{figure}

Figure \ref{fig_cycles}(a) shows the latitude-time variation of the
mean (azimuthally averaged) toroidal magnetic field at a depth near the
bottom of the CZ. The mean toroidal magnetic field tends to
be of opposite signs for the two hemispheres, and exhibits an irregular
cyclic behavior with oscillations of the field
strength on time scales ranging from about 5 years to about 15 years and  
undergoes irregular sign/polarity reversals.
The strongest mean toroidal field is concentrated near the bottom of the
CZ (see Figure \ref{fig_cycles}(b)), peaking at about $7$ kG.
Figure \ref{fig_cycles}(c) shows a shell-slice of $B_{\phi}$ at a depth
near the bottom of the CZ, at a cycle maximum phase indicated
by the green line in Figure \ref{fig_cycles}(a). It shows that strong
toroidal fields $B_{\phi}$ of a preferred sign (opposite
for the two hemispheres) are concentrated in individual channels or filaments 
in each hemisphere, reaching peak field strength of about $30$ kG, which
exceeds the field strength in equipartition with the local r.m.s
convective speed ($B_{\rm eq} \approx 13$ kG) but is below the 
equipartition field strength corresponding to the peak down flow speed
($B_{\rm eq} \approx 63$ kG).
Thus these strong field filaments are not passively advected by convective
flows but would be pinned down by the strong down flows if in their paths.

\subsection{Maintenance of the solar-like differential rotation \label{subsec:diffrot}}

\begin{figure}
\includegraphics[width=0.175\textwidth]{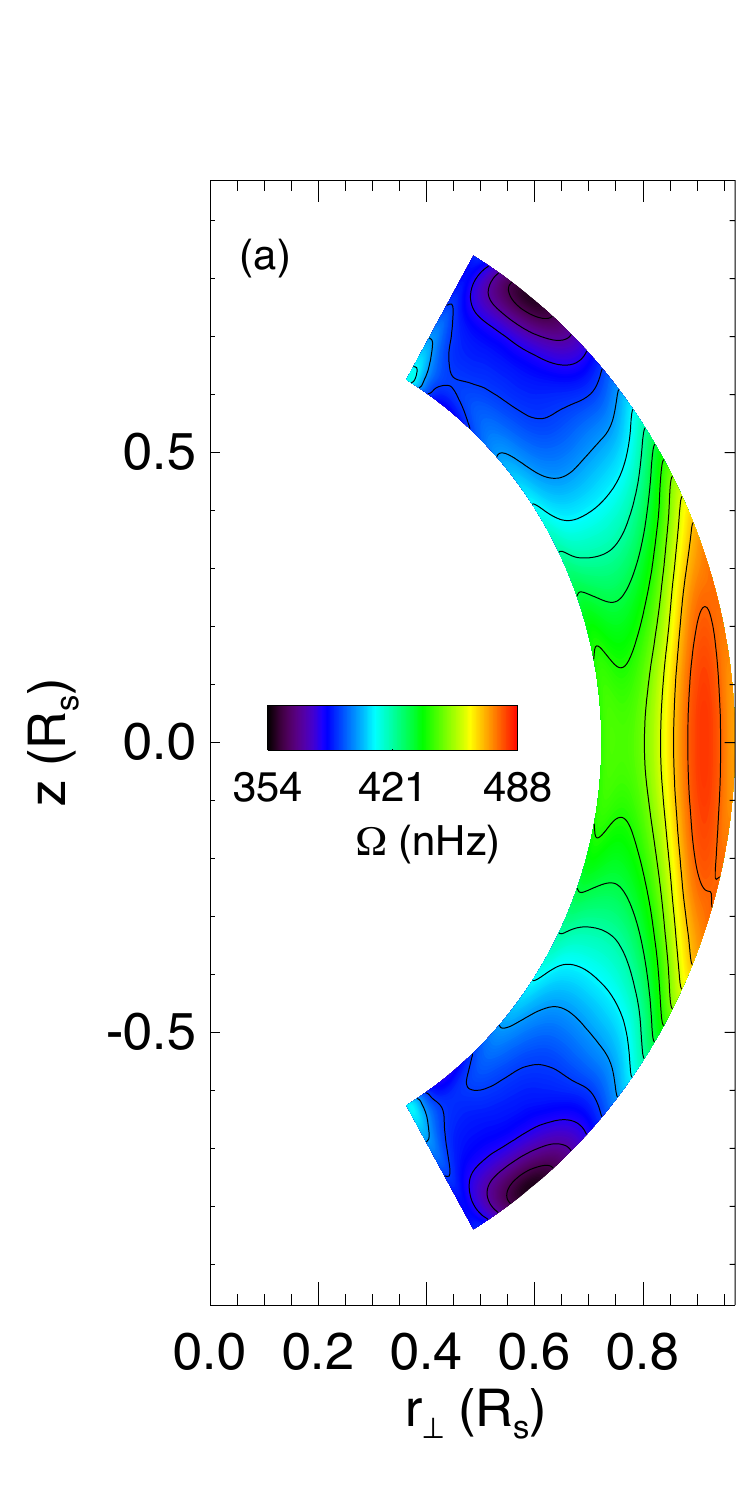}
\includegraphics[width=0.175\textwidth]{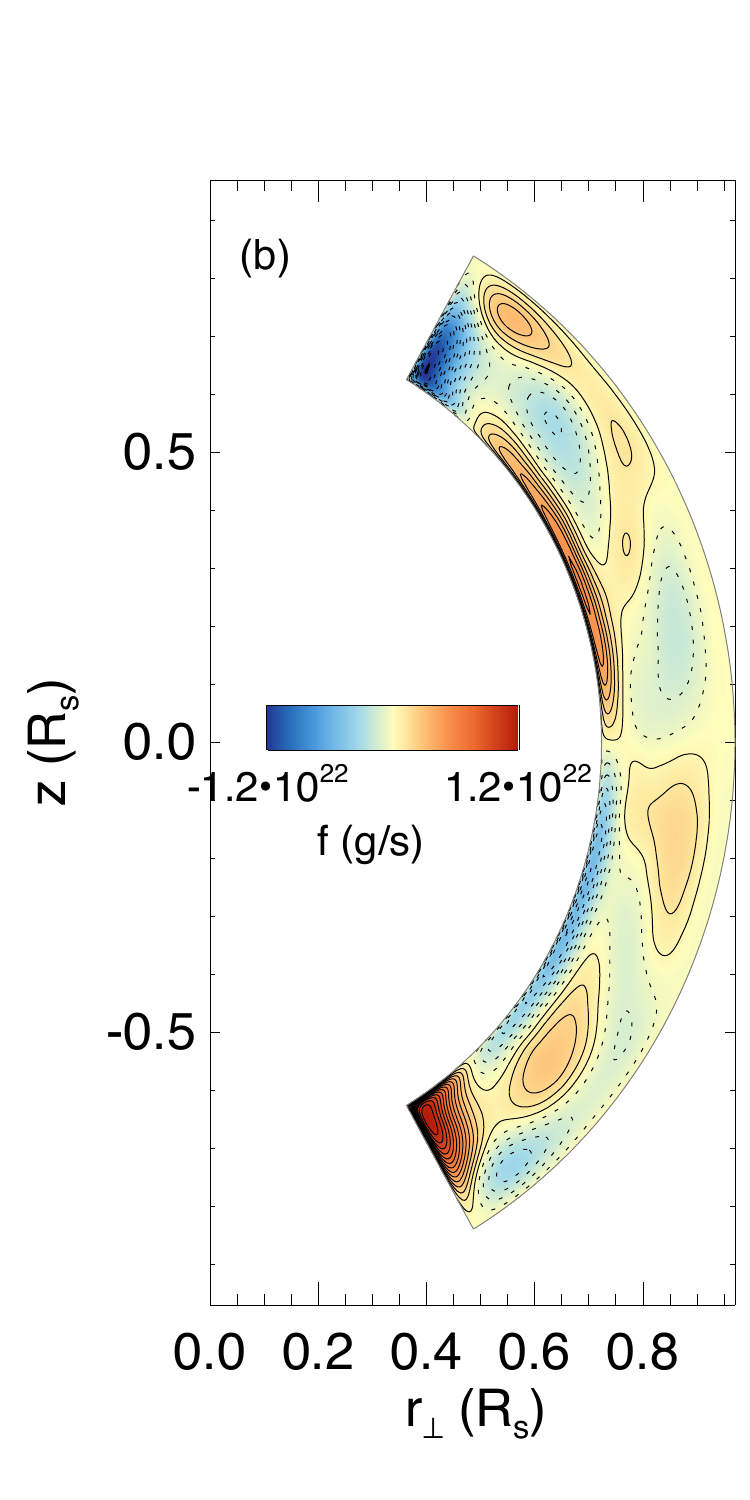}
\includegraphics[width=0.175\textwidth]{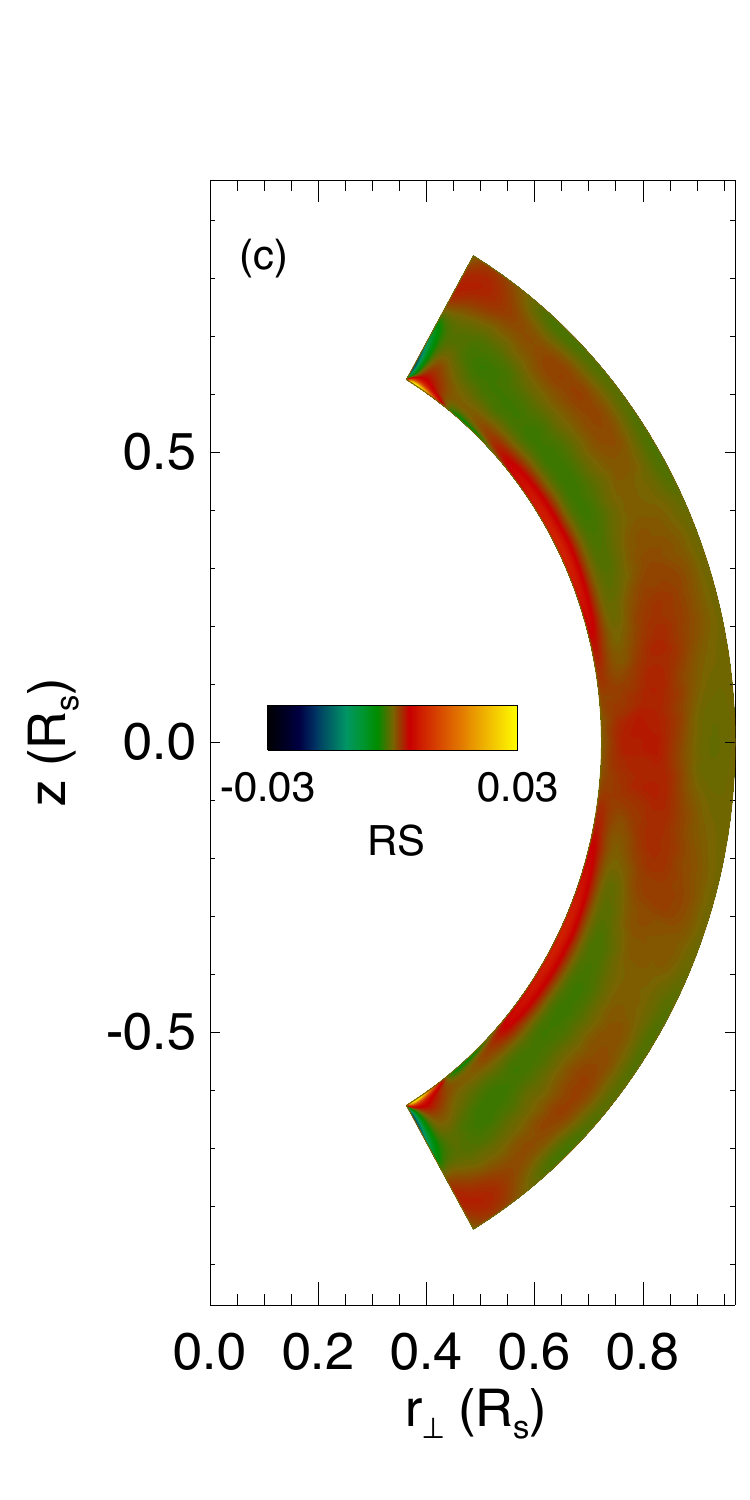}
\includegraphics[width=0.175\textwidth]{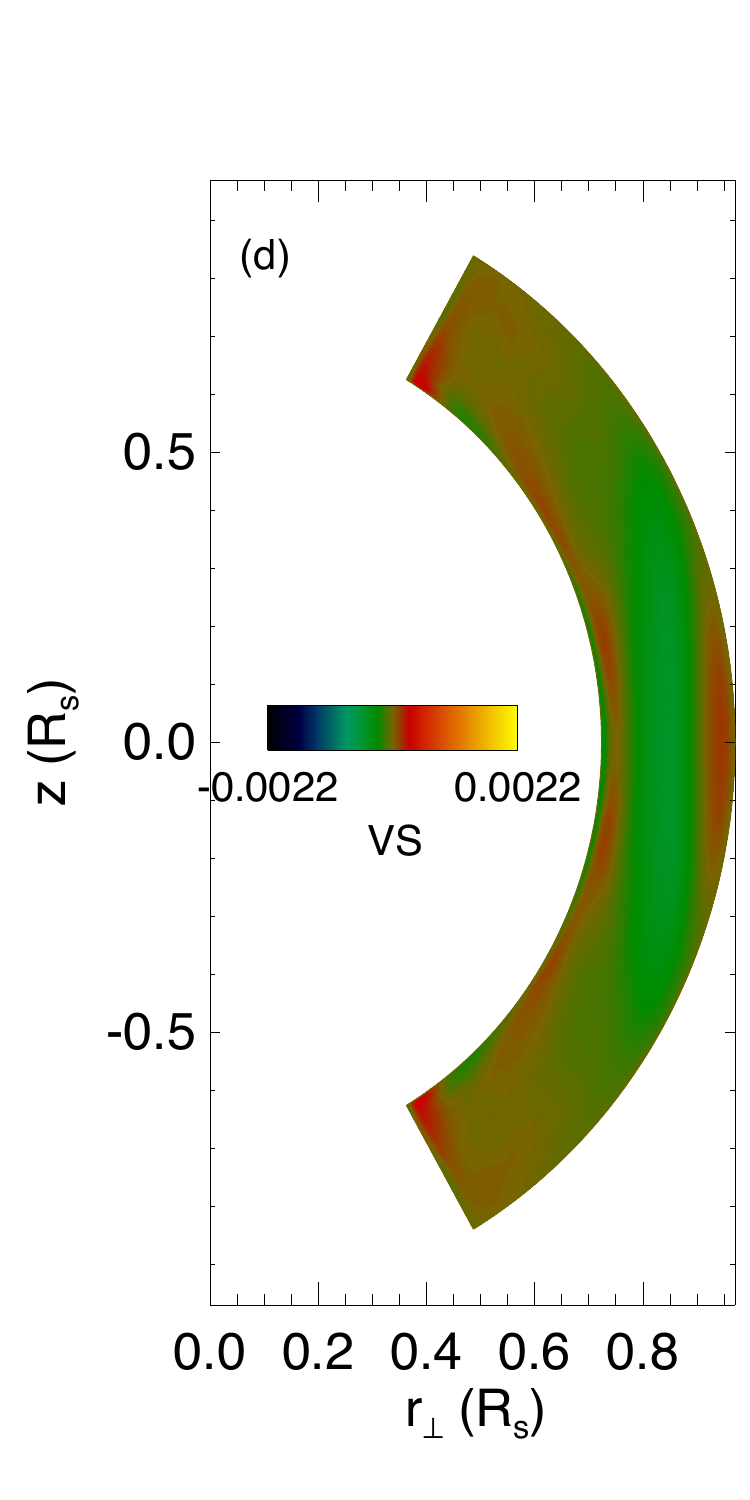}
\includegraphics[width=0.175\textwidth]{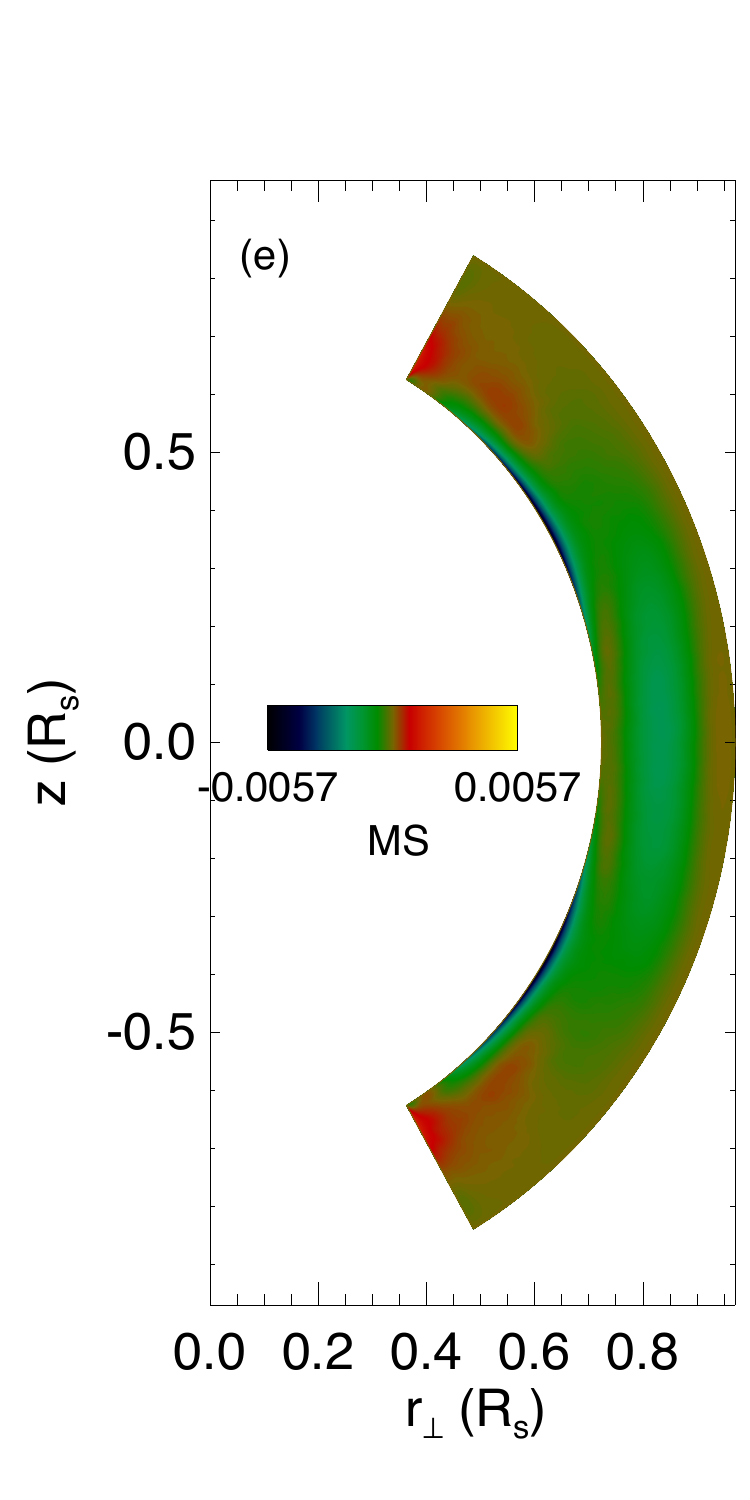} \\
\includegraphics[width=0.175\textwidth]{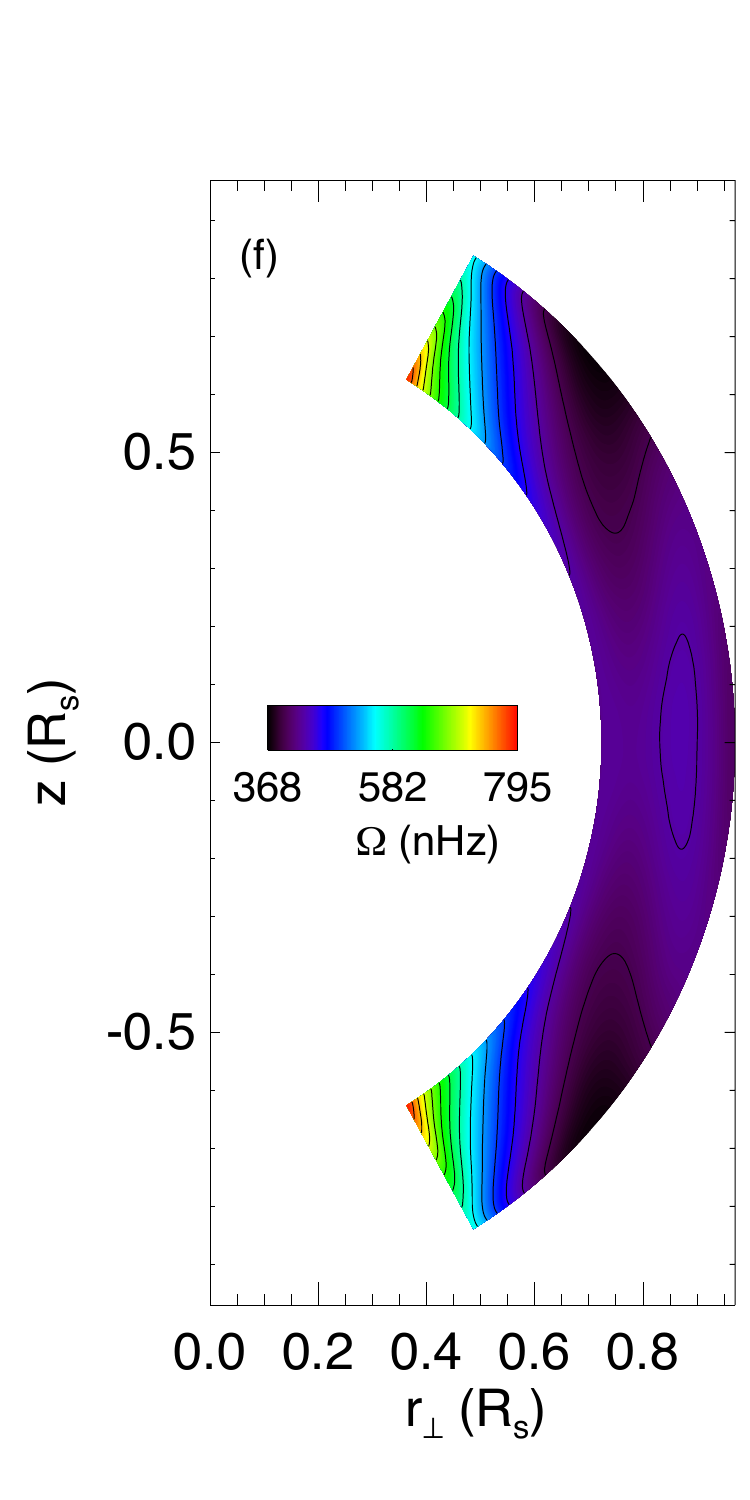}
\includegraphics[width=0.175\textwidth]{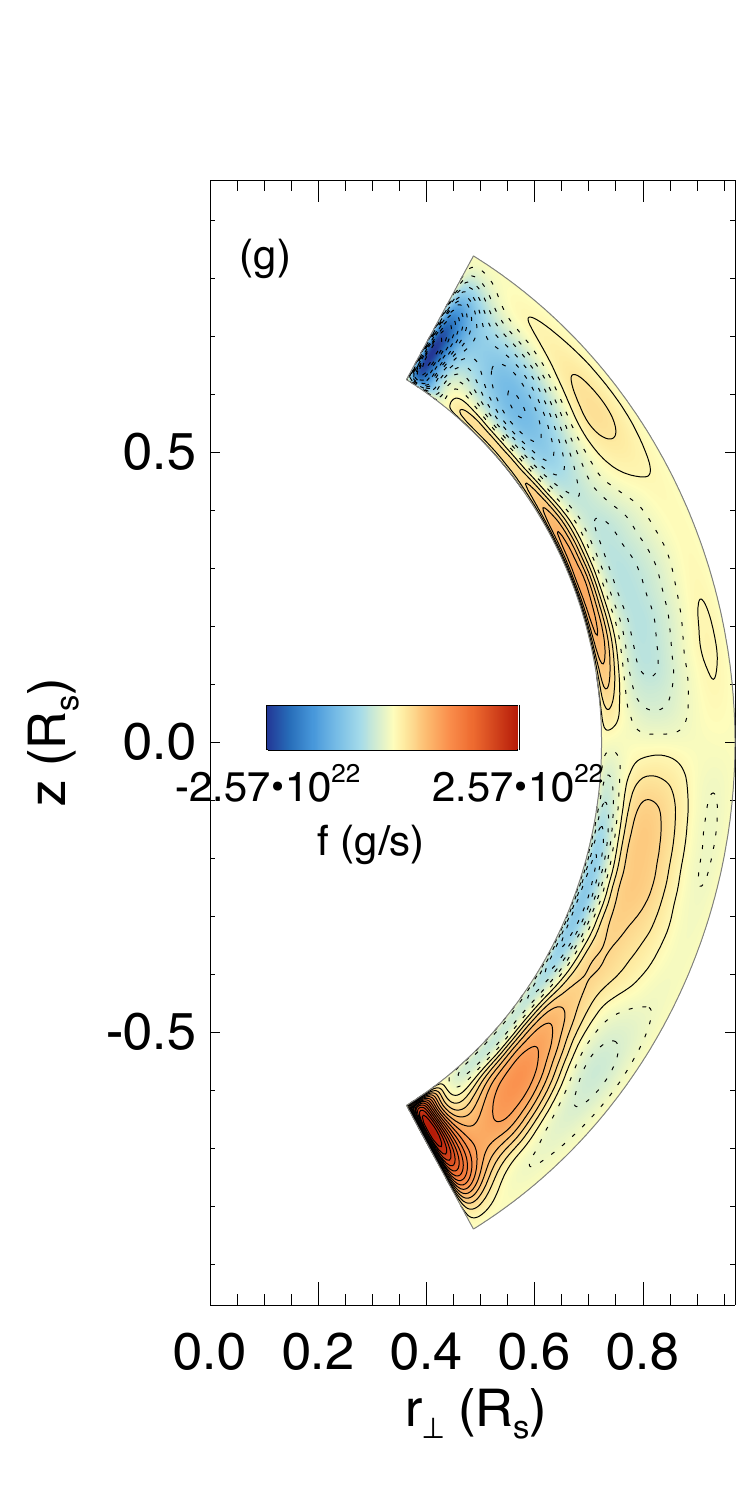}
\includegraphics[width=0.175\textwidth]{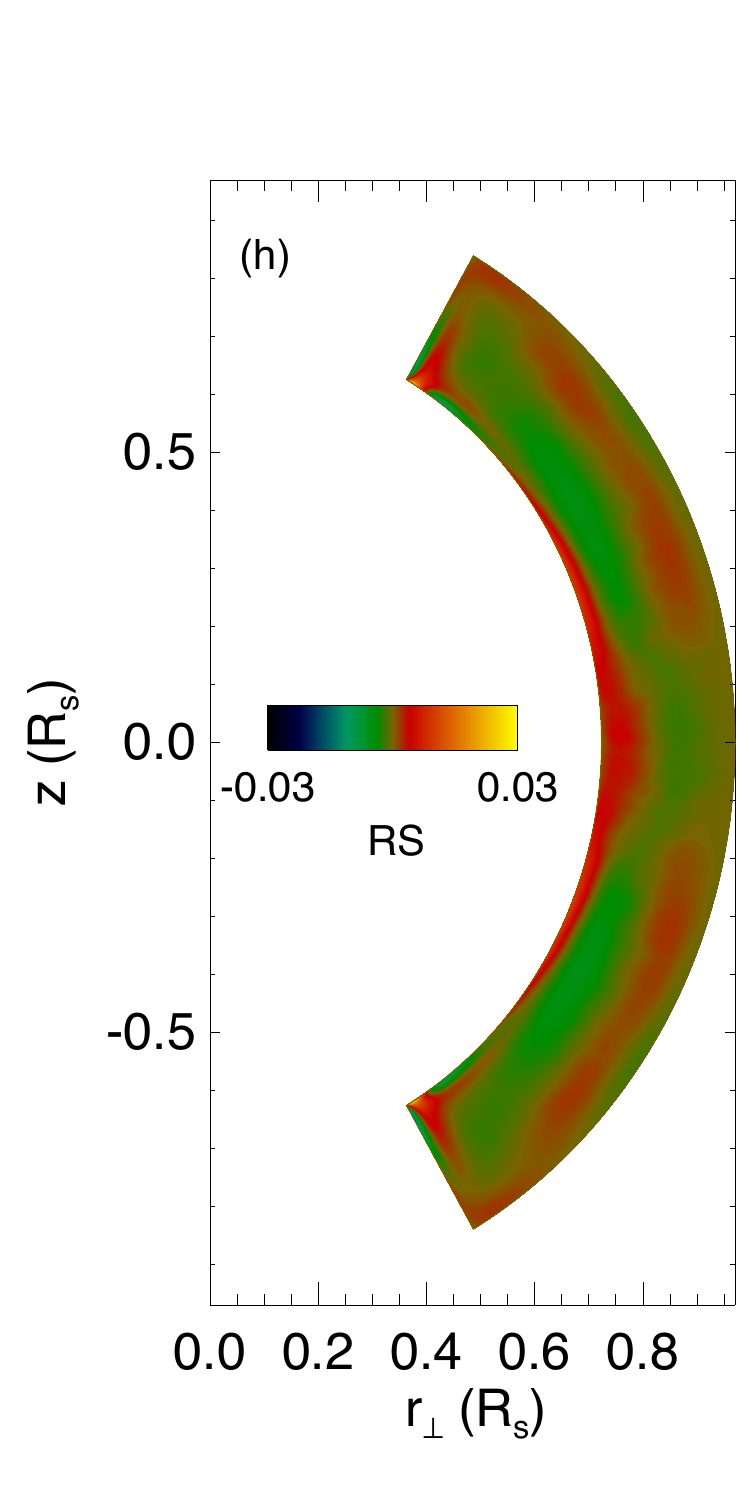}
\includegraphics[width=0.175\textwidth]{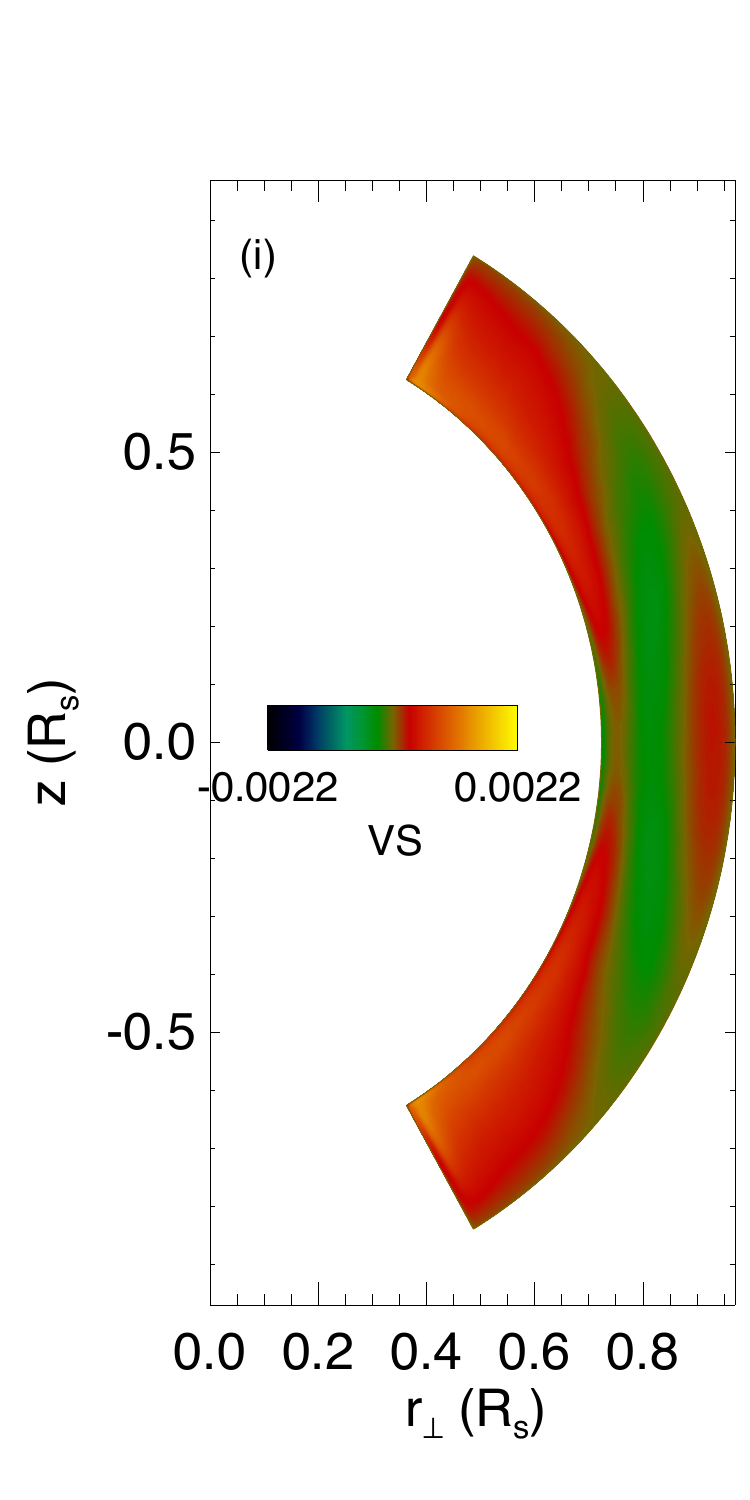} \\
\includegraphics[width=0.175\textwidth]{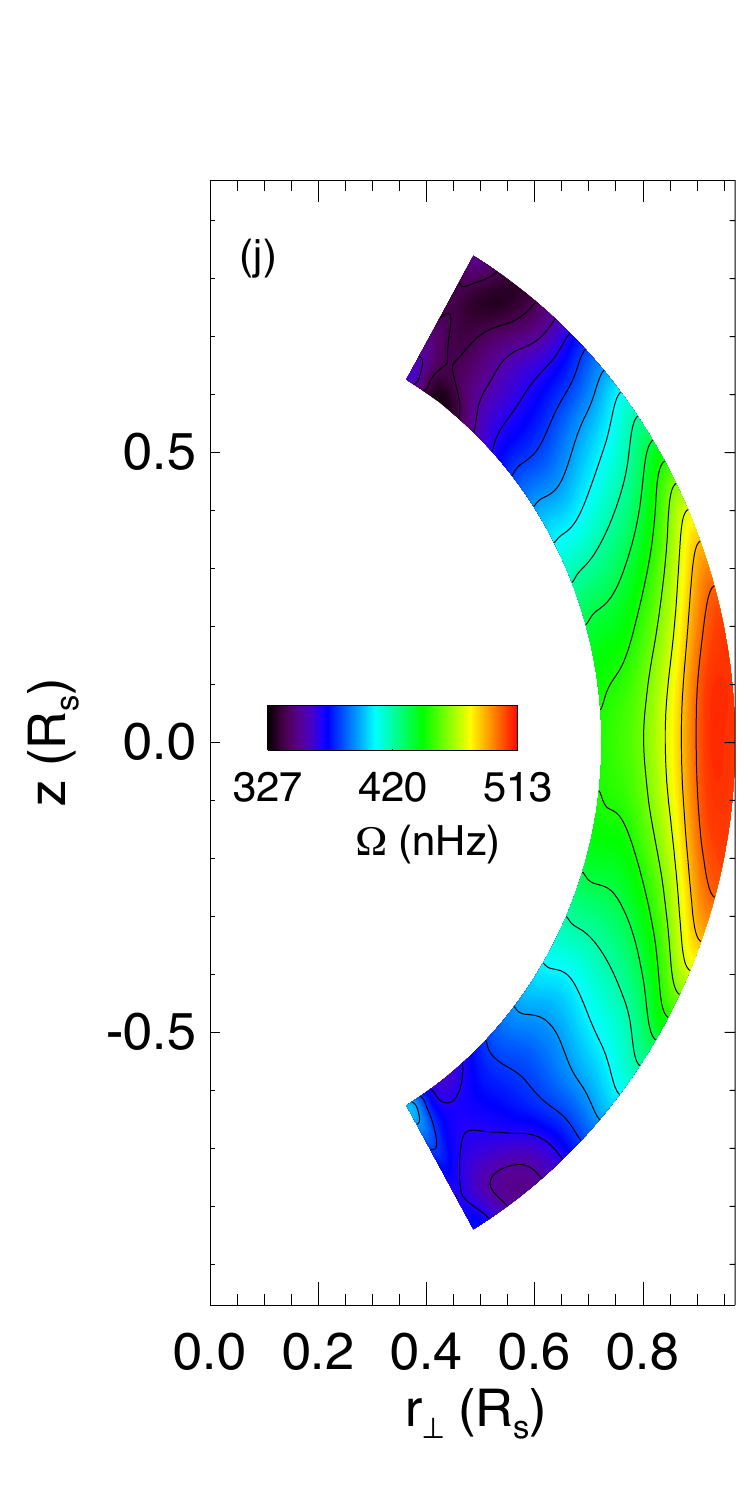}
\includegraphics[width=0.175\textwidth]{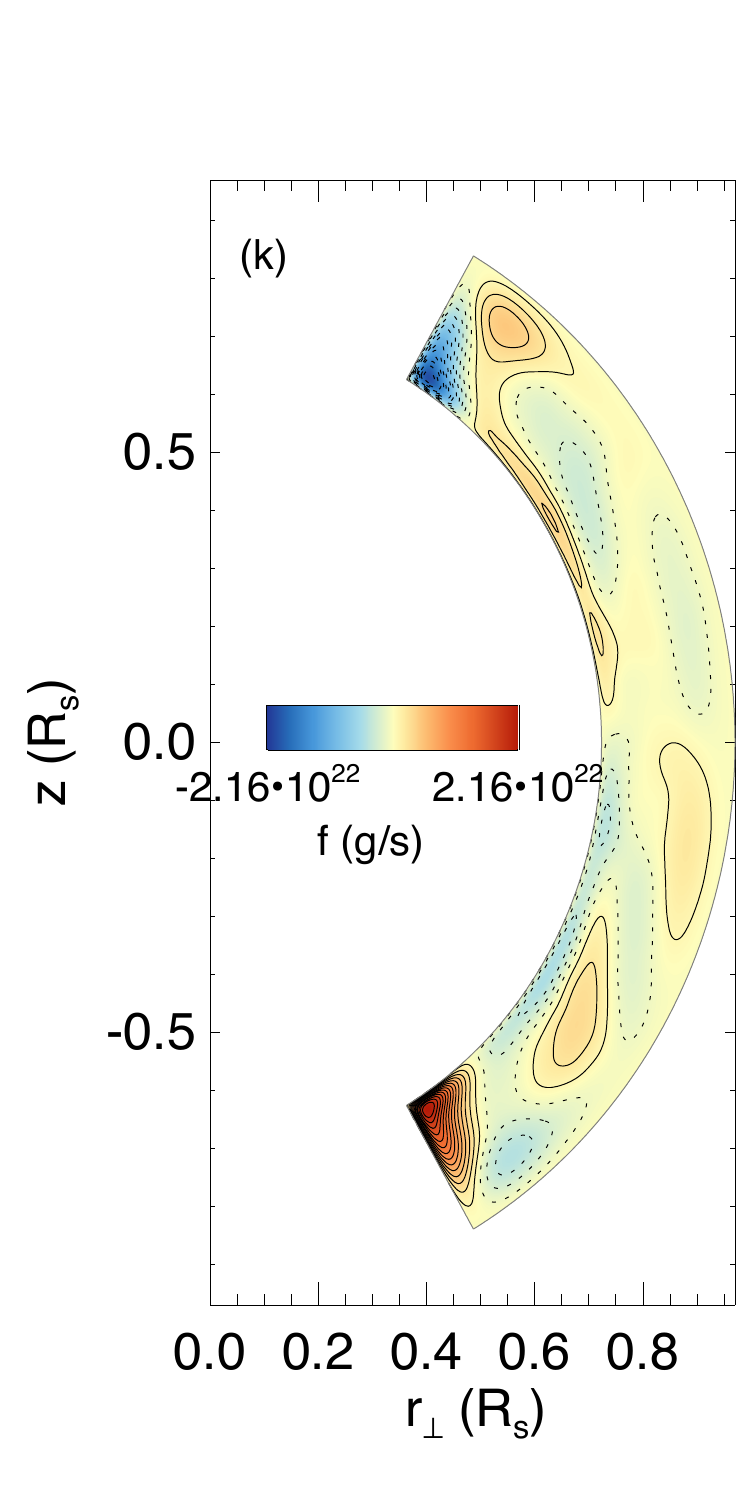}
\includegraphics[width=0.175\textwidth]{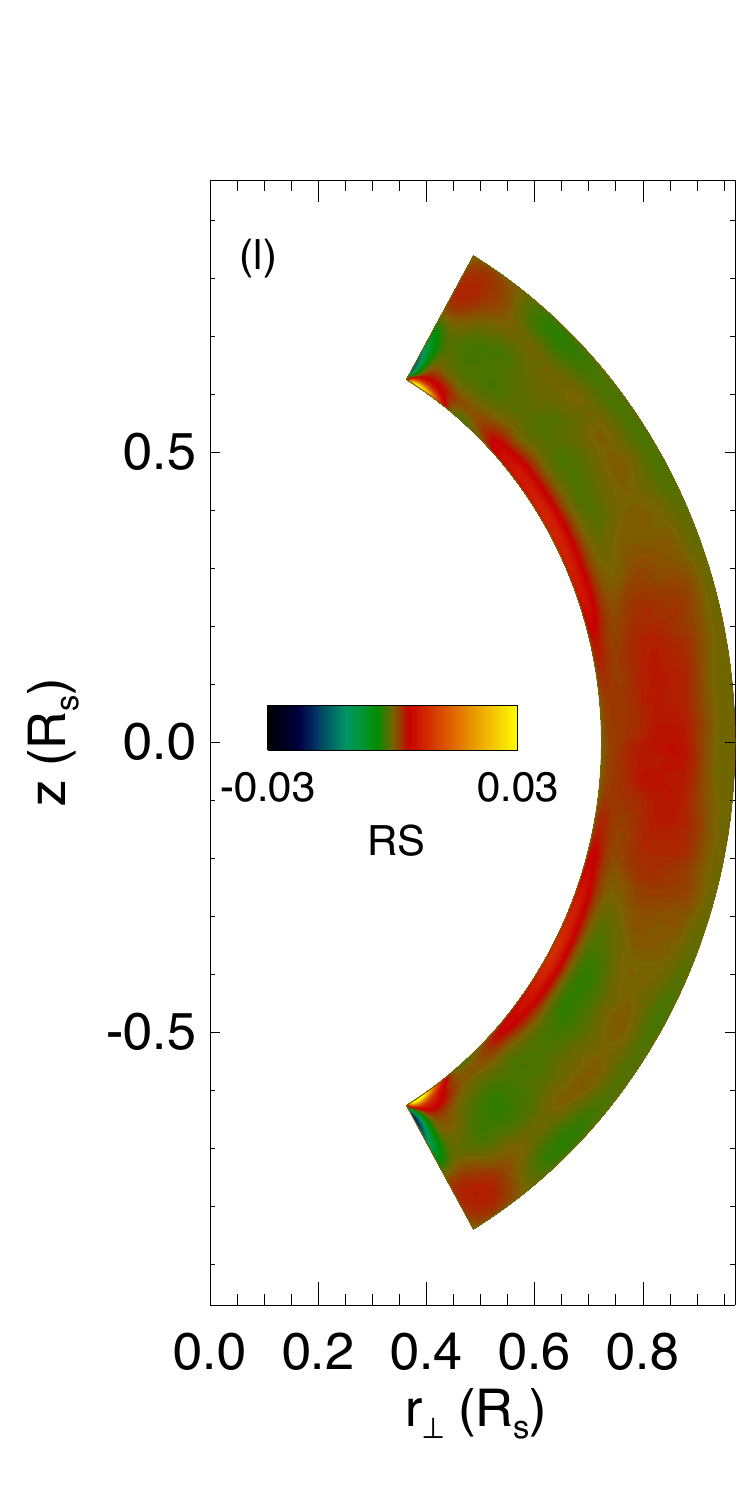}
\includegraphics[width=0.175\textwidth]{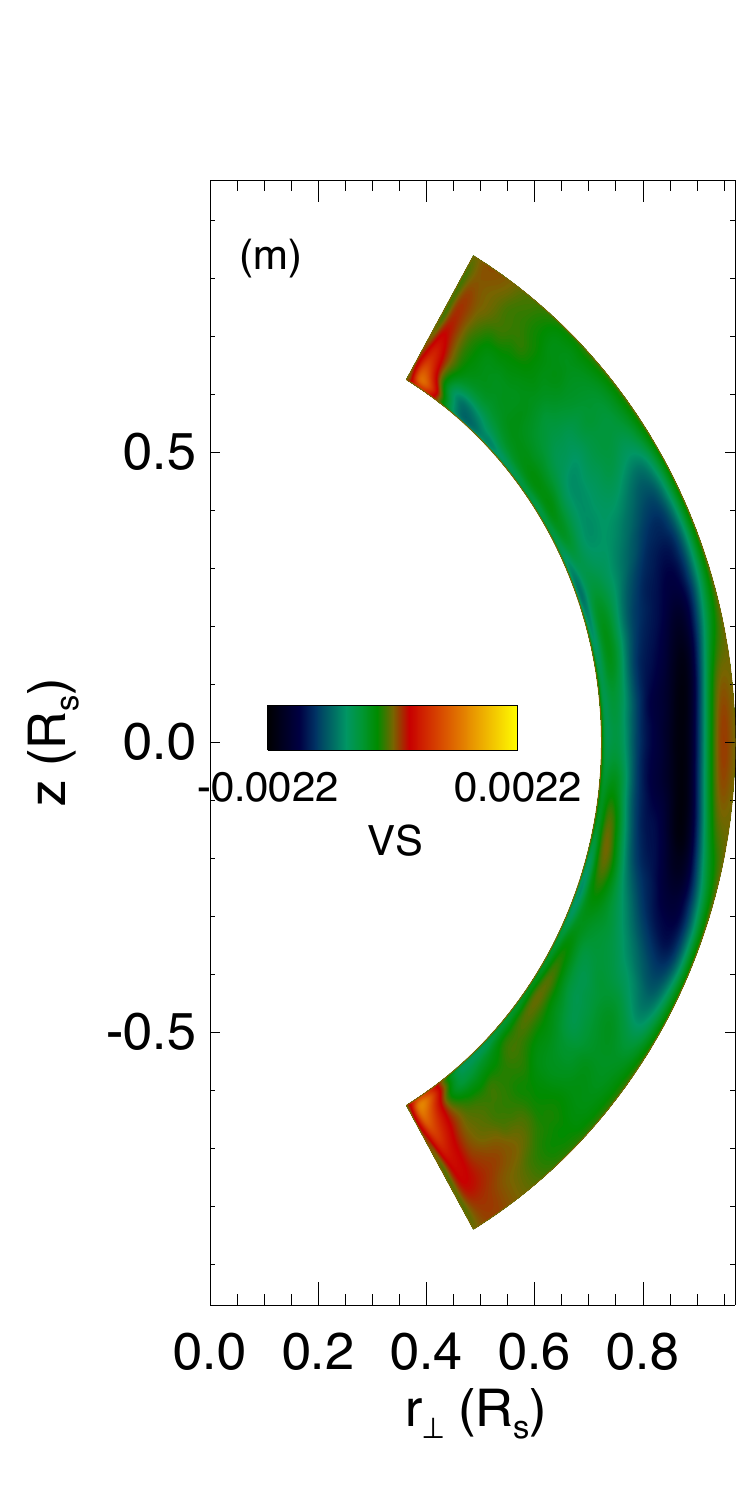}
\caption{Top row panels from left to right show the time and azimuthally averaged mean
meridional profile of angular velocity (a), meridional flow mass flux (b), angular momentum flux density in the $r_{\perp}$ direction due to the Reynolds stress (c), the viscous stress (d), and the Maxwell stress (e) resulting from the dynamo simulation. Middle row panels from left to right show the same as those of the top row except for the results from the corresponding HD simulation and there is not a panel for the Maxwell stress. Bottom row panels from left to right show the same as the middle row, except for the results from the HVHD simulation.  See text for the expressions for the various angular momentum flux density RS, VS, and MS.}
\label{fig_diffrot}
\end{figure}

Figure \ref{fig_diffrot}(a) shows the time and azimuthally averaged rotation
rate in the convective envelope self-consistently maintained in the
convective dynamo simulation.  It shows a solar-like differential rotation
profile \citep[e.g.][]{Thompson:etal:2003} with a faster
rotation rate at the equator than at the polar region by about 30\% of the
mean rotation rate, and more conical shaped iso-rotation contours in the
mid latitude zone. The time and azimuthally averaged mean meridional flow
pattern is shown in Figure \ref{fig_diffrot}{b} in terms of the mass flux function
$f$ where $\rho_0 < \! \! {\bf v} \! \!> =
\nabla \times [ ( f / r \sin \theta ) {\hat {\bf \phi}} ]$.
The meridional circulation has a complex
multi-cell structure with a counter-clockwise (clockwise) cell pattern in
the low latitude region of the northern (southern) hemisphere,
i.e. a poleward near-surface flow in the low latitude region. 

Interestingly, we find that the presence of the magnetic field is necessary
for the self-consistent
maintenance of the solar-like differential rotation profile in the current
parameter regime.  We have
carried out a hydrodynamic simulation (hereafter referred to as the HD case)
which is identical to the present convective dynamo
simulation except that the magnetic field is set to zero, and found that a
very different differential rotation profile (Figure \ref{fig_diffrot}(f)) is
established in the statistical steady state of the HD simulation.
It shows a significantly larger differential rotation with a faster rotation
rate in the polar region than at mid-latitudes and the equator.  The
iso-rotation contours are also more cylindrical.  The meridional flow (Figure
\ref{fig_diffrot}(g)) shows
a more prominent counter-clockwise (clockwise) cell pattern in the northern
(southern) hemisphere in the mid depths in the CZ, with much weaker
reversed cells in the near surface layer.

\begin{figure}
\centering
\includegraphics[width=0.45\textwidth]{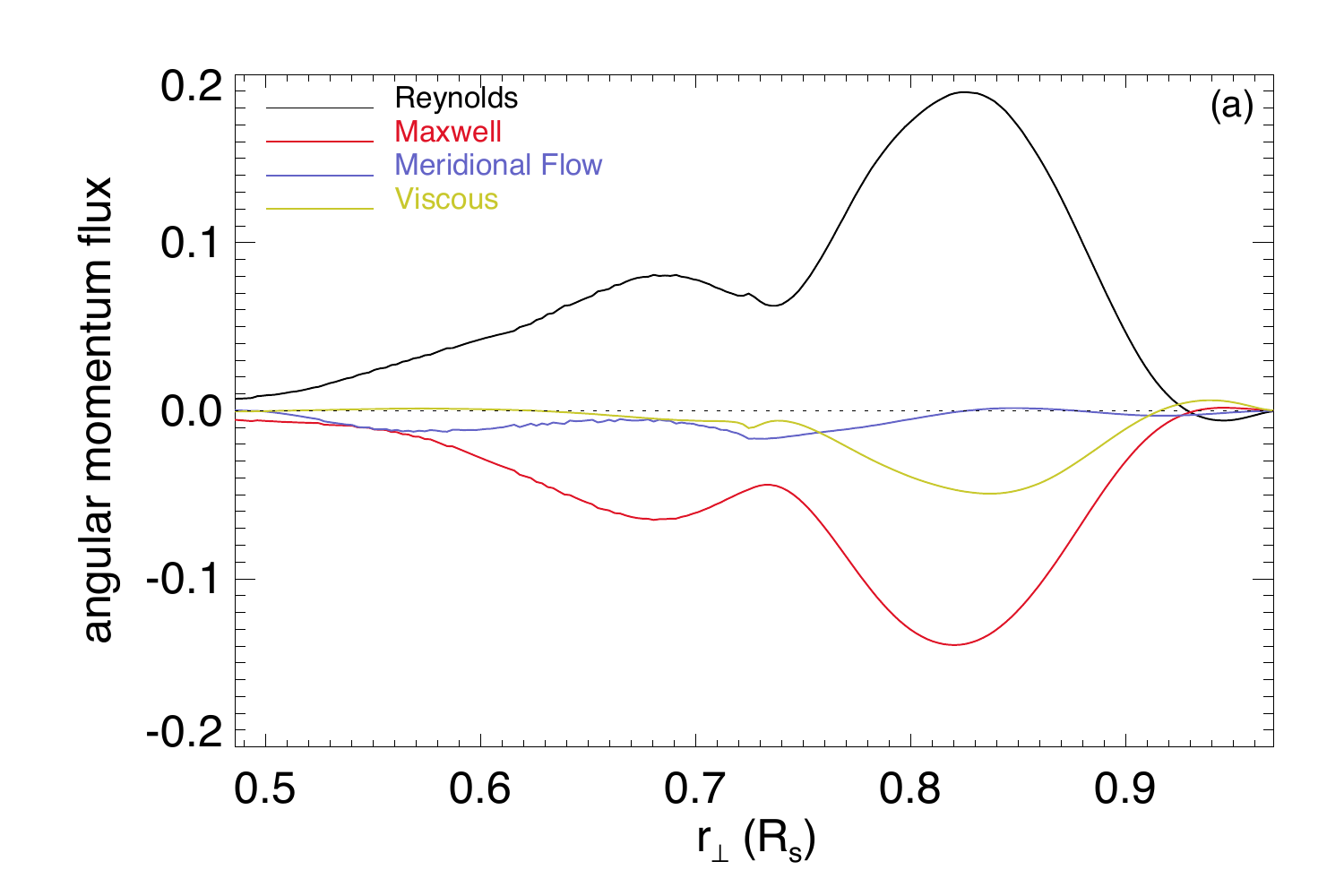} \\
\includegraphics[width=0.45\textwidth]{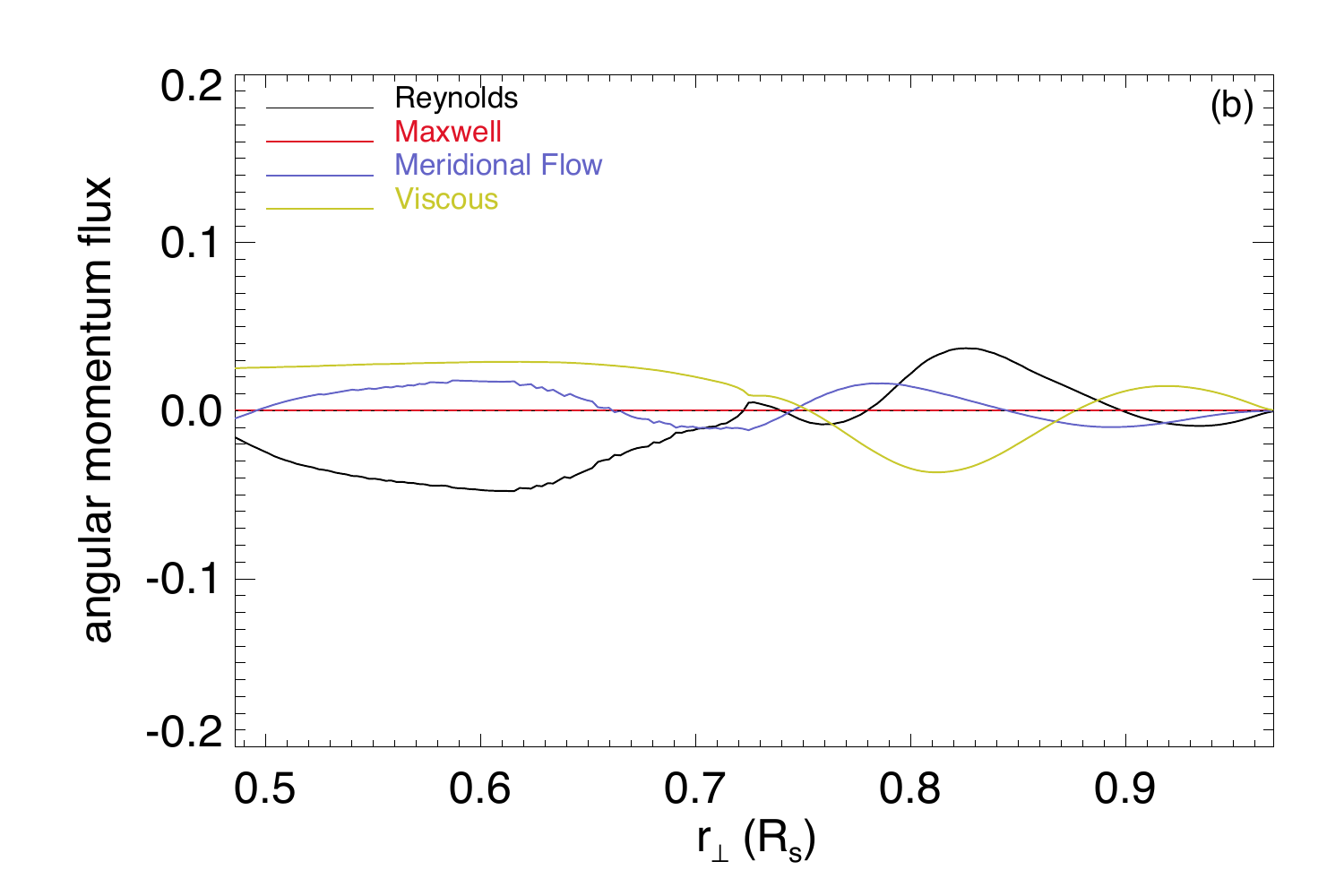} \\
\includegraphics[width=0.45\textwidth]{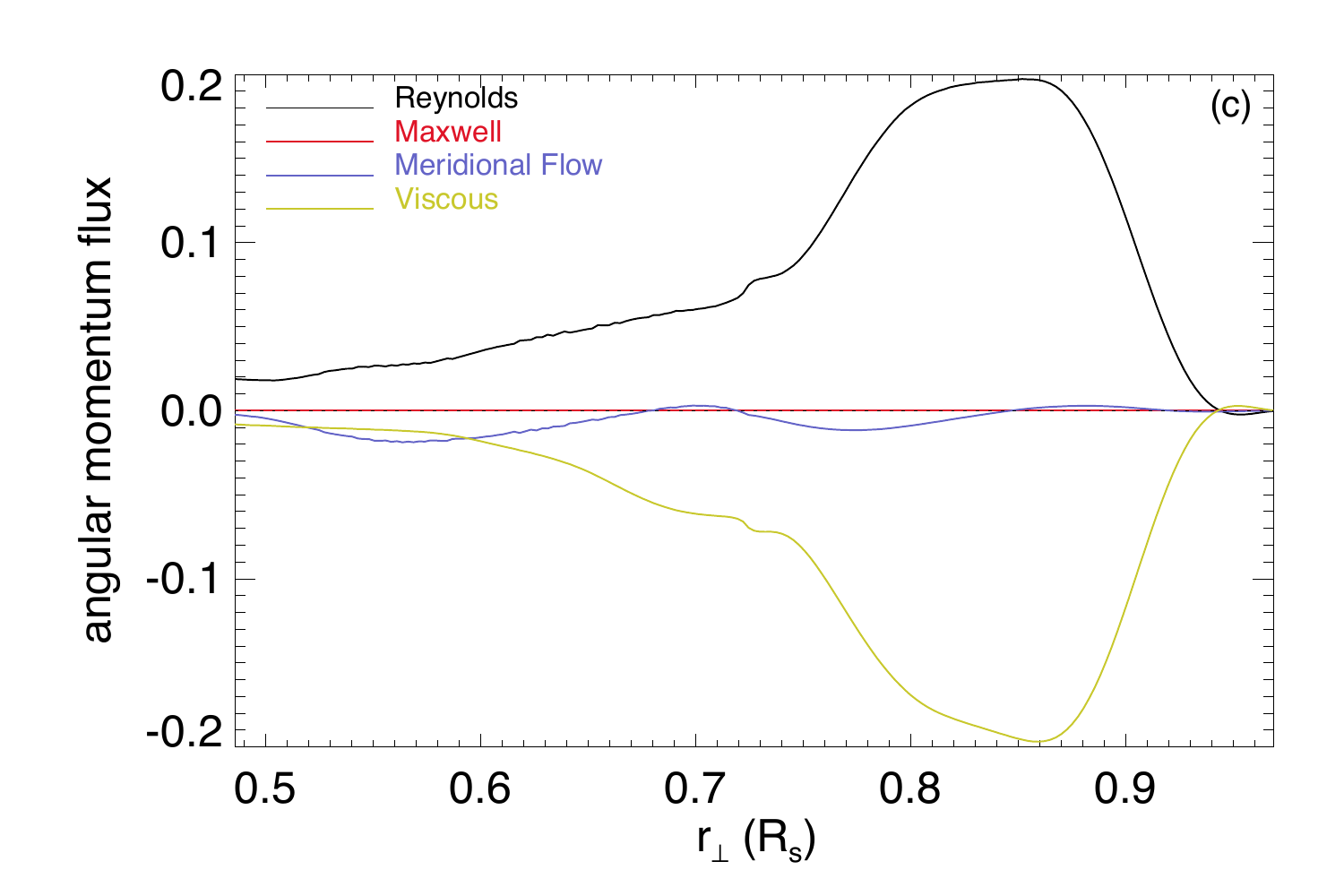}
\caption{Integrated angular momentum fluxes across cylinders centered on the rotation axis as
a function of the radial distance $r_{\perp}$ from the rotation axis for respectively
the dynamo case (a), the corresponding HD case (b), and the HVHD case (c). The
$r_{\perp}$ of the tangent cylinder of the base of the CZ is $0.722 R_s$ }
\label{fig_angularmomflux}
\end{figure}

To compare the transport of angular momentum in the dynamo and the HD cases,
we show in Figures \ref{fig_diffrot}(c), \ref{fig_diffrot}(d), and \ref{fig_diffrot}(e)
the meridional profile of the angular momentum flux density
in the $r_{\perp}$ direction (perpendicular to and away from the rotational
axis) due respectively to the Reynolds stress of the rotationally influenced
convection (panel (c)):
\begin{equation}
RS = \rho_0 r_{\perp} < \! \! v'_{r_{\perp}} v'_{\phi} \! \!>,
\label{eq:RS}
\end{equation}
the viscous stress (panel (d)):
\begin{equation}
VS = \rho_0 \nu r_{\perp} ( <\!\!S_{\phi r } \!\!> \sin \theta
+ <\!\!S_{\phi \theta} \!\!> \cos \theta ),
\label{eq:VS}
\end{equation}
and the Maxwell stress (panel (e)):
\begin{equation}
MS=- r_{\perp} \frac{1}{4 \pi} < \!\! B_{\phi} B_{r_{\perp}} \!\! >,
\label{eq:MS}
\end{equation}
where $< >$ denotes time and azimuthal averages and $'$ denotes the azimuthally
varying component.
The meridional profiles of the angular momentum flux density $RS$ and $VS$ for
the corresponding HD case
are shown in Figures \ref{fig_diffrot}(f) and \ref{fig_diffrot}(g).
We find that there is a significant difference in the angular momentum
flux density by the Reynolds stress
between the dynamo and HD cases.
The $RS$ for the dynamo case shows an overall more outward transport in
its meridional distribution compared to that for the corresponding HD case.
Near the lower boundary in the mid latitude range, the presence of
the concentrated magnetic field (which helps to damp the convective downflows)
results in a more enhanced outward $RS$ flux at the lower boundary layer.
The angular momentum flux density $MS$ due to the Maxwell stress in the
dynamo case (Figure \ref{fig_diffrot}(e)) is found to oppose $RS$, and its
strength is the greatest at the lower boundary layer where the magnetic
field concentrates.
The angular momentum flux density $VS$ simply acts to reduce the differential
rotation as expected for both the dynamo and the HD cases.
The difference between the dynamo and the HD cases is more clearly seen
by evaluating the net angular momentum fluxes in the $r_{\perp}$ direction
integrated over individual concentric cylinders of radii $r_{\perp}$ centered
on the rotation axis, as shown in Figure \ref{fig_angularmomflux}(a) for the
dynamo case and Figure \ref{fig_angularmomflux}(b) for the corresponding
HD case.
It can be seen that the net angular momentum flux due to $RS$ is outward
throughout (except near the top boundary) for the dynamo simulation (black
curve in Figure \ref{fig_angularmomflux}(a)),
which drives a faster rotation in the outer equatorial region.
The net angular momentum flux due to $RS$ is mainly counteracted by the
net flux due to $MS$ by the Maxwell stress, with the remaining difference
balanced by the significantly smaller net fluxes due to $VS$ and the meridional
flow.  In contrast, the HD simulation shows a significant inward angular
momentum transport due to the Reynolds stress (black curve in
Figure \ref{fig_angularmomflux}(b)) across the inner cylinders in the high
to mid latitude region. This drives a faster rotation in the polar region.
Thus it appears that the presence of the
magnetic field alters the convective flows such that the resulting
Reynolds stress from the convective motions
produces a more outward (away from the rotation axis)
net transport of the angular momentum needed to drive a solar-like
differential rotation.
The Rossby number $R_{O} = v_{\rm rms,all} / (\Omega H_p )$ is about 0.74 for
the dynamo case and 0.96 for the HD case, where $v_{\rm rms,all} = 125 $ m/s
for the dynamo case and $157 $ m/s for the HD case, is the
r.m.s. velocity (with the azimuthally averaged mean flow velocity taken out)
averaged over the entire volume, and $H_p$ is the pressure scale height at
the bottom of the convection zone.  The Rossby number measures the importance
of the Coriolis force in the force balance. The lower Rossby number
in the dynamo simulation shows that the the magnetic fields suppress the
convective motions so that they are more rotationally constrained.

A recent systematic study by \citet{Gastine:etal:2014} of rotating stellar
convection considering a wide range of models shows that the differential
rotation profile transitions from being solar-like, with a faster rotating
equator, to being anti-solar, with a faster polar rotation rate, at a value
of about 1 for the Rossby number. This result is found to be quite general,
independent of the detailed model setup (presence of a
magnetic field, thickness of the convective layer, density stratification).
Our HD case with $R_O = 0.96$ appears to be very close to the transition,
and a reduction of $\sim 23$\% of the overall r.m.s. velocity and
$R_O$ by the presence of the magnetic field in the dynamo case is able to significantly
alters the angular momentum transport by the rotationally constrained
convection, leading to a transition into the solar-like differential
rotation.
We note that even though the HD case is quite close to the transition, its
anti-solar differential rotation appears to be a stable solution not
dependent on the history, i.e. not one of two bistable states
\citep{Gastine:etal:2014}. We have arrived at the statistically steady HD
solution with the anti-solar differential rotation by either
starting from the initial setup with a seed velocity field as described in
section \ref{sec:model},
or by starting from the statistically
steady state dynamo solution (for which the differential rotation is
solar-like) and zero out the magnetic field.
The solar-like differential rotation obtained in the convective dynamo
simulation also appears to be a stable solution that is not dependent on the history.  
The differential rotation at the pole and equator remain statistically
steady without exhibiting any systematic drift for the $ \gsim 74 $ year
period (comparable to the maximum viscous time scale near the bottom
of CZ) we have run after the dynamo solution has reached a statistical steady
state.  Also as we start the dynamo simulation from the
initial setup described in section \ref{sec:model}, we find that the
convective dynamo goes through an earlier phase of anti-solar
differential rotation (for $\sim 6$ years) before it evolves towards 
the solar-like differential rotation profile as the mean entropy gradient 
and the convective energy flux settle down to their statistical steady state.
Thus it appears that the solar-like differential rotation is the preferred
stable solution in the dynamo case.

The transition to a solar-like differential rotation can
alternatively be achieved in the non-magnetic hydro simulations
by simply increase the viscosity to reduce $R_O$.
We have run another hydrodynamic simulation (hereafter referred to
as the HVHD case, meaning ``high viscosity hydro'') where we increase the
viscosity $\nu$ by 5 times (with the same ${\rho_0}^{-1/2}$ depth
dependence) compared to the HD case (or the dynamo case).
The resulting r.m.s. velocity of the statistical steady convection reached
is $v_{\rm rms,all}=120 m/s$ and the Rossby number $R_O = 0.71$, much closer
to those of the dynamo case.
The bottom row panels of Figure \ref{fig_diffrot} show respectively the
resulting differential rotation profile (Figure \ref{fig_diffrot}(j)),
meridional circulation (Figure \ref{fig_diffrot}(k)),
the angular momentum flux density in the $r_{\perp}$ direction,
$RS$ (Figure \ref{fig_diffrot}(l)) due to the Reynolds stress,
and $VS$ (Figure \ref{fig_diffrot}(m)) due to the viscous stress.
The integrated net (outward) angular momentum fluxes across concentric
cylinders of radius $r_{\perp}$ centered on the rotation axis is shown
in Figure \ref{fig_angularmomflux}(c).
It is found that a solar-like differential rotation profile
(Figure \ref{fig_diffrot}(j)) with faster
rotating equator and with a more conical iso-rotation contours in
mid-latitude zones is established, although the contrast of rotation
rate between the equator and the polar region is bigger, about $44$\% of
the mean rotation rate (compared to about $32$\% in the dynamo case).
The mean meridional circulation
(Figure \ref{fig_diffrot}(k)) shows a
counter-clockwise (clockwise) cell pattern in
the low latitude region of the northern (southern) hemisphere,
i.e. a poleward near-surface flow in the low latitude region, similar
to the dynamo case (Figure \ref{fig_diffrot}(b)).
We find that the angular momentum transport in the $r_{\perp}$
direction due to the Reynolds stress for the HVHD case
is very similar to that for the dynamo case, both in the meridional
profile of the flux density
(Figure \ref{fig_diffrot}(l) compared to Figure \ref{fig_diffrot}(c))
as well as in the integrated net flux across the constant
$r_{\perp}$ concentric cylinders (Figure \ref{fig_angularmomflux}(c)
compared to Figure\ref{fig_angularmomflux}(a)).
But this similar outward net angular momentum flux by the Reynolds
stress is now balanced almost entirely by the transport due to the
viscous stress, with the absence of the Maxwell stress which is
the major component that balances the angular momentum flux by
the Reynolds stress in the dynamo case.
The comparison between the dynamo and the HVHD cases
suggests an effective role of enhanced viscosity played by the
magnetic fields, which (1) suppresses the large scale convective
motions such that they are more rotationally constrained (lower $R_O$) to
produce an outward transport of the angular momentum by the Reynolds stress,  
necessary to drive a solar-like differential rotation, and (2) takes up the main role
to balance the Reynolds stress transport with the Maxwell stress
instead of the viscous stress.

Further in Figure \ref{fig_heatfluxes_hdandhdhvis} we show the various
horizontally integrated energy fluxes through the domain for the
HD case (upper panel) and the HVHD case (lower panel),
in comparison with the energy fluxes shown
in Figure \ref{fig_heatfluxes} for the dynamo case.
It can be seen that the dynamo case and the HVHD case show
a similar convective energy flux $L_{\rm conv}$ (reaching about 66\% $L_s$
in the dynamo case and about 60\% $L_s$ in the HVHD case).
In both the dynamo and the HVHD cases, the
downward kinetic energy flux $L_{\rm kin}$ (reaching about 16\% $L_s$ in the
dynamo and 10\% $L_s$ in the HVHD case) is significantly
reduced compared to the HD case (reaching about 45\% $L_s$).
In fact in the HD case (upper panel in Figure
\ref{fig_heatfluxes_hdandhdhvis}), the downward kinetic energy flux is
so large that the outward convective energy flux $L_{\rm conv}$ exceeds
the solar luminosity in the middle of the convection zone to counter it. 
The result here indicate again the similar role played by the magnetic fields
and the enhanced viscosity in suppressing the downward convective flows.

\subsection{Emerging flux \label{subsec:emgflux}}

\begin{figure}
\centering
\includegraphics[width=0.5\textwidth]{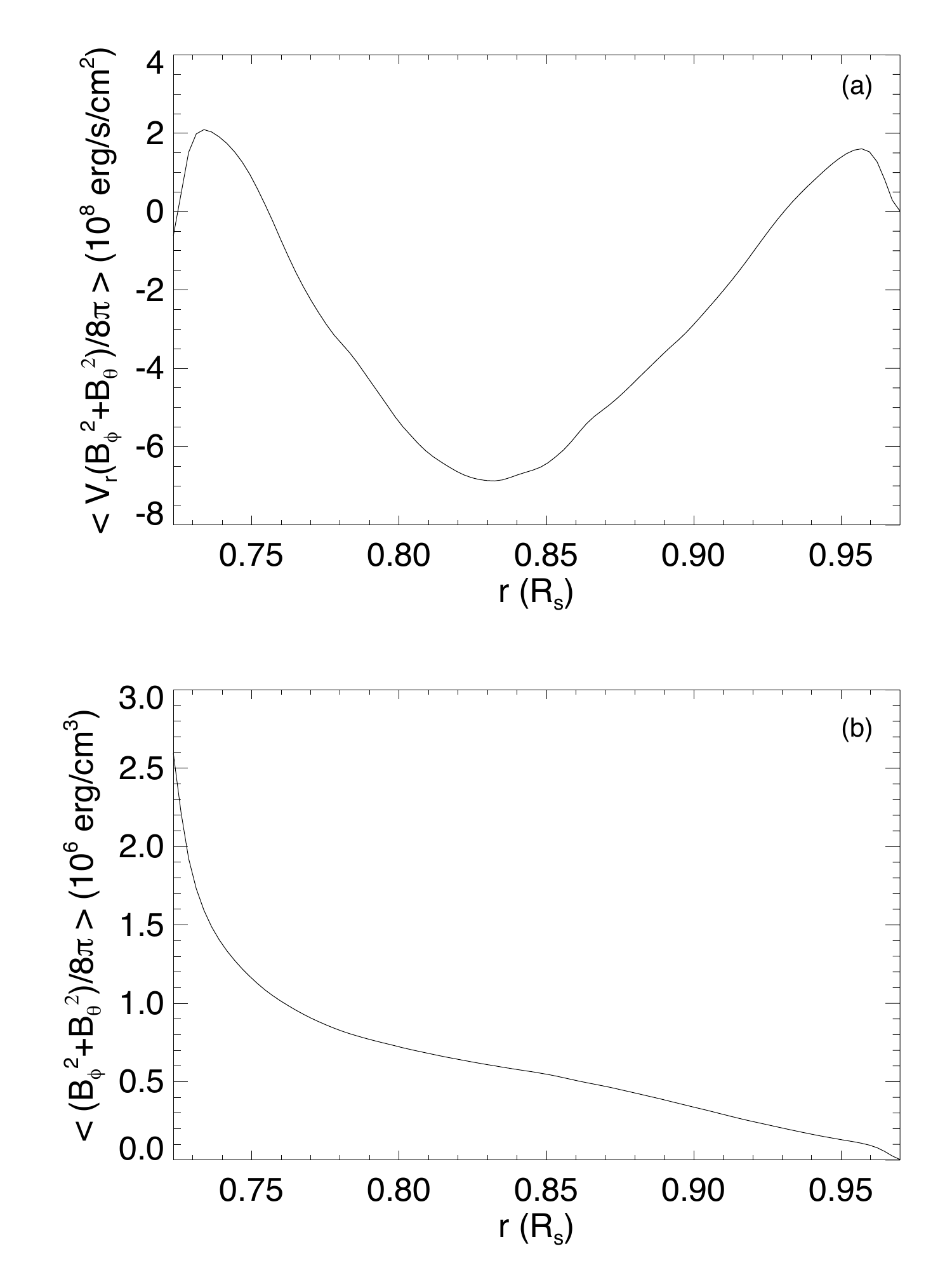}
\caption{(a) Advective flux of the horizontal magnetic field energy,
and (b) the distribution of horizontal magnetic field energy as a function
of depth.}
\label{fig_emtransport}
\end{figure}

In the convective dynamo, the large-scale mean toroidal field as shown in
Figure \ref{fig_cycles}(b) is produced by the latitudinal differential
rotation shearing a dipolar poloidal mean field. The reason that the mean
toroidal field is concentrated towards the bottom of the CZ is mainly due to
a downward advective transport of the magnetic energy in the bulk of the
convection zone, as represented by
$< \!\! v_r ({B_{\theta}}^2+{B_{\phi}}^2) / 8 \pi \!\!>$ shown
in Figure \ref{fig_emtransport}(a).
This causes the distribution of the magnetic energy (for both the
mean field and the small scale field) to be strongly concentrated
towards the bottom (see Figure \ref{fig_emtransport}(b)).
The decrease with depth of the magnetic diffusivity $\eta$ would
also promote stronger fields towards the bottom but is less important
here because of the small magnitude of $\eta$
and the long diffusive time scale: $(\Delta r)^2 / \eta \sim 10$
years, using the peak $\eta$ value near the surface and the depth
of the CZ domain $\Delta r$.
The advective time scale for the downward magnetic energy transport
across the convection zone is $\Delta r / u_m \sim 0.5 $ year is significantly
shorter, where we have used $u_m = < \!\! v_r ({B_{\theta}}^2+{B_{\phi}}^2) \!\!> /
< \!\! {B_{\theta}}^2+{B_{\phi}}^2 \!\!>$ evaluated at the middle of
the convection zone as a measure of the transport speed. Thus the advective
transport acts more quickly.

In the midst of magneto-convectoin, we find occasional active region
like flux emergence events in the top layer of the simulation domain.
Such an example is shown in Figure \ref{fig_emgevent}, where panels
(a), (b), (c), and (d) show respectively snapshots of $B_r$,
$B_\phi$, $v_r$ and $v_\phi$ at a constant $r$ slice at the depth
of $30$ Mm below the photosphere.
The location of the emerging bipolar region is indicated by an arrow
in the panels. It is characterized by a diverging bipolar pattern
in $B_r$ (panel (a)) and the emergence of a strong toroidal field patch
reaching a peak field strength of $9800$ G (panel (b)) (see also the online
movie). The emerging region corresponds to an up flow region in $v_r$
(panel (c)), but the upward velocity is not significantly different from that
of other up flow convective cells.  The zonal velocity $v_{\phi}$ of the
emerging region shows a diverging pattern, and when averaged over the
emerging region, is $\sim 100$ m/s
faster than the mean zonal velocity of that latitude.
Figure \ref{fig_emgevent}(e) shows the subsurface 3D magnetic field
configuration in the convective envelope by showing field lines traced from
randomly seeded points throughout the volume.  The field lines are colored
based on their azimuthal field $B_{\phi}$ as indicated by the color table.
It can be seen that relatively more coherent bundles of strong toroidal
flux are embedded in the
turbulent magnetic fields. In Figure \ref{fig_emgevent}(f), regions of strong
field strength where the Alfv\'en speed ($v_a$) exceeds the r.m.s. convective
velocity ($v_{\rm rms}$) for the corresponding depth is outlined with the
equipartition iso-surfaces (with $v_a / v_{\rm rms} = 1$),
which are again colored based on the $B_{\phi}$ value on the iso-surfaces.
There is a systematic preference for these strong flux regions to be green
or of negative $B_{\phi}$ (red or of positive $B_{\phi}$) in the
northern (southern) hemisphere.
The arrows in Figures \ref{fig_emgevent}(e) and \ref{fig_emgevent}(f)
mark the toroidal flux bundle with super-equipartition field strength that gives
rise to the emerging region.

Figure \ref{fig_emgzoomview} shows a more zoomed in view of the thermodynamic
properties of the emerging region at the same depth as that shown in the
upper 4 panels of Figure \ref{fig_emgevent}.
We see that there is a systematic reduction of density
(i.e. buoyant, see Figure \ref{fig_emgzoomview}(b)) and pressure
(Figure \ref{fig_emgzoomview}(d)) in the
emerging region compared to the surrounding, although the reduction
magnitude is rather moderate compared to the fluctuations seen in strong
downflow lanes and in strong vertical flux tubes in the downflow lanes.
The temperature change in the emerging region compared to the surrounding
is smaller, partly due to the large thermal conduction.
Averaged over the emerging region (area enclosed in the yellow contour in
Figure \ref{fig_emgzoomview}(a), $< \!\! \rho_1 / \rho_0 \!\! > \approx -1.6
\times 10^{-5}$, $< \!\! p_1 / p_0 \!\! > \approx - 1.1 \times 10^{-5}$, and
$< \! \! T_1 / T_0 \! \! > \approx 0.5 \times 10^{-5}$.  It can be seen
that the temperature change is relatively small compared to the density and
pressure change in the emerging region.  This suggests that the buoyancy or
reduction in density is mainly due to the reduction in gas pressure provided
largely by the presence of the magnetic pressure, instead of mainly due to
an increase in temperature.  In other words, the buoyancy contribution is
more from the magnetic buoyancy than the thermal buoyancy.

As can be seen from Figure \ref{fig_emgevent}(e), the emerging flux region is
the apex of a roughly east-west oriented (toroidal) super-equipartition flux bundle which
remain relatively coherent for some distance, before the two ends connect in
complex ways to other flux systems.  The following end of the coherent flux
bundle extends into the middle of the CZ.  The fact that the emerging flux
has a prograde zonal speed of $\sim 100$ m/s relative to the mean zonal speed
of the latitude indicates that it is not a toroidal flux tube rising in
isolation from the bottom of the CZ. Because if it were it would have a
retrograde flow due to angular momentum conservation as is
found in many previous studies of isolated rising flux tubes in the rotating
solar CZ \citep[e.g.][]{Caligari:etal:1995,Fan:2008,Fan:2009:lrsp}.
The emerging flux bundle must have well mixed with the local plasma through
reconnections, and is continually sheared and amplified by the differential
rotation and the local flows against resistive dissipation.
The sequence of images in Figure \ref{fig_shearhairpin} show the sub-surface
development of the super-equipartition emerging flux bundle (marked by
the arrow) over a 9 day period prior to the time of the flux emergence event
shown in Figure \ref{fig_emgevent}. It shows that local shear in the upper convection
zone contributes significantly to the development of the emerging flux bundle.
The left column images
show iso-volumes of super-equipartition fields with the surface
of the volume colored by $B_{\phi}$. The right column images show
representative field lines traced from the iso-volume corresponding
to the emerging flux bundle.
It can be seen that a segment of a super-equipartition flux bundle in the
middle of the convection zone is sheared and stretched in the prograde
direction into a hairpin turn with the upper side
of the hairpin forming the emerging flux bundle reaching the top boundary.

\begin{figure}
\includegraphics[width=0.5\textwidth]{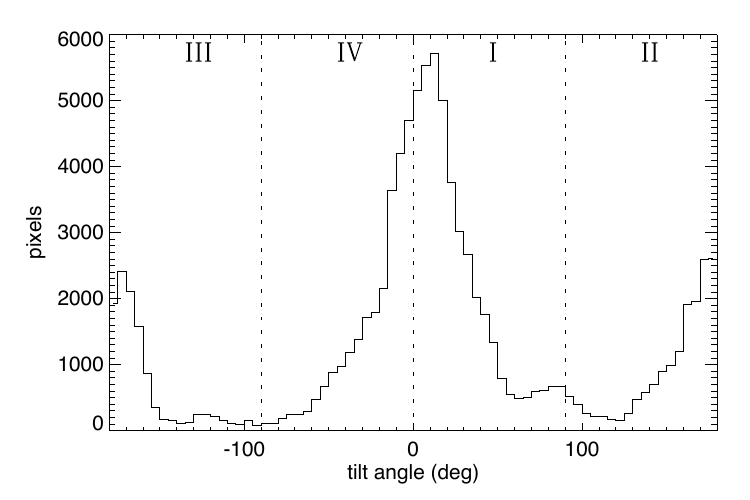}
\caption{Distribution of the tilt angles of the horizontal field
vectors in strong emerging field areas at 30 Mm depth. See text about the
tilt angle quadrants I, II, III, and IV.}
\label{fig_tiltdistr}
\end{figure}

We have done a statistical study of the super-equipartition emerging fields.
For a time period of about 1 year centered at the cycle maximum phase
(green line in Figure \ref{fig_cycles}(a)) and at an interval of 12 hours,
we find in
the shell slice at 30 Mm depth all the area where the emerging horizontal field
exceeds $\sqrt{2}$ times the field strength in equipartition with the r.m.s.
convective velocity of that depth.
For each pixel (or grid point) of the selected emerging field area, we compute
a tilt angle of the horizontal field vector based on the local $B_{\phi}$ and
$B_{\theta}$. The resulting tilt angle distribution of all the pixels
is shown in Figure \ref{fig_tiltdistr}.  The quadrant of the tilt angle
is such that, if the sign of the azimuthal field $B_{\phi}$ is
consistent with Hale's polarity rule of the cycle,
i.e. negative (positive) in the northern (southern)
hemisphere, then the tilt angle falls in quadrants I and IV.
If the horizontal field vector is tilted clockwise (anti-clockwise) from
the cycle preferred azimuthal field direction in the northern (southern)
hemisphere by an acute angle, consistent with the mean
tilt of solar active regions, then the tilt angle falls in quadrant I.
We find from Figure \ref{fig_tiltdistr} that there is a preference for
Hale's polarity rule for the emerging azimuthal field by a ratio of 2.4 to 1
in the area. For those pixels satisfying Hale's rule the mean tilt angle
is $7.5^\circ$, with an estimated uncertainty of $1.6^\circ$.
Thus the super-equipartition emerging
fields have a statistically significant mean tilt similar
to the active region mean tilt \citep[e.g.][]{Stenflo:Kosovichev:2012}.

\section{Discussions\label{sec:conc}}
We have presented a 3D convective dynamo driven by the solar radiative
diffusive heat flux in the solar CZ, and with a
latitudinal gradient of entropy imposed at the bottom, representing the
tachocline induced thermal variation that can break the Taylor-Proudman
constraint in the CZ \citep{Rempel:2005}.  The convective dynamo produces a large
scale mean field that undergoes irregular cycles and polarity reversals,
and self-consistently maintains a solar-like differential rotation with
faster rotation at the equator than at the polar region by about 30\% and
more conical iso-rotation contours in mid latitudes
\citep[e.g.][]{Thompson:etal:2003}.

The irregular cyclic behavior of the mean field in our model differs from
those in the literature \citep[e.g.][]{Ghizaru:etal:2010, Racine:etal:2011,
Kapyla:etal:2012}.
By comparing the Reynolds number $R_e$ and the Coriolis number $C_O$
(as defined in \citet{Kapyla:etal:2012}) achieved in our dynamo run with
those of the cyclic dynamo simulation presented in
\citet{Kapyla:etal:2012} (see section \ref{subsec:overview}),
we find that their convective dynamo is operating
in a significantly more rotationally constrained regime, with their $C_O$
being about 5 times ours. Our convective flows appear to be only moderately
more turbulent compared to theirs as reflected in the similar order of
magnitude for the $R_e$ values.  If we consider our convective
r.m.s. velocity to be similar to theirs, then their dynamo would be
effectively operating in a stellar envelope that is rotating at about 5 times
the solar rotation rate.
This is probably the main reason we obtain a very different mean field
dynamo behavior compared to that of \citet{Kapyla:etal:2012}.
Our irregularly cycling convective dynamo model also differs significantly
in many ways from the cyclic convective dynamo model described in 
\citet{Ghizaru:etal:2010, Racine:etal:2011}.
Their dynamo model used an implicit large eddy code with no
explicit viscosity, magnetic diffusion, and thermal diffusion. But our
convective dynamo appears to be operating in a more turbulent regime
if we compare the convective downflow speed obtained:
about $25$ m/s at $r=0.954 R_s$ in \citep{Ghizaru:etal:2010} vs.
our $\sim 300$ m/s at the same depth.
Given that both models use the solar rotation rate for the convective
envelope, this suggests that their dynamo model is also operating in a
significantly more rotationally constrained regime with a significantly
lower Rossby number compared to ours.
The Newtonian cooling treatment of the entropy equation
used in \citet{Ghizaru:etal:2010, Racine:etal:2011} is very different
from our treatment of the energy transport which forces the solar luminosity
through the convective domain.
Furthermore, their model includes a sub-adiabatically stratified overshoot
layer at the bottom of the convection zone which our model does not have. 
All these contribute to the significant differences in the resulting dynamo
behavior.

In both \citet{Kapyla:etal:2012} and \citet{Racine:etal:2011}, because of
the significantly higher $C_O$ or lower $R_O$, the convective flows
in their corresponding hydro cases in the absence of the magnetic fields
are already driving
a solar-like differential rotation, and the addition of the magnetic fields
in their dynamo cases appear to mainly reduce the differential rotation
\citep{Charbonneau:2013}.
On the other hand for our convective dynamo in a significantly less
rotationally constrained regime with $R_O$ closer to 1, the presence of the
magnetic field is found to play an important role for the maintenance of the
solar-like differential rotation, without which a faster rotating polar
region results, as is shown with the corresponding HD simulation.
A solar-like differential rotation profile can alternatively be achieved
in the hydro case by increasing the viscosity as shown in the HVHD simulation.
The comparison between the dynamo case and the HVHD case indicate that in
several aspects the magnetic field plays an effective role of enhanced
viscosity to (1) suppress the large scale convective motions such that they
become more rotationally constrained to produce an outward transport of
angular momentum by the Reynolds stress needed to drive a solar-like
differential rotation, (2) take up the main role of balancing the Reynolds
stress transport with the Maxwell stress transport instead of the viscous
stress under low viscosity conditions, and (3) reduce the downward kinetic
energy energy flux.
Our resulting differential rotation self-consistently maintained 
in the convective dynamo simulation is in fairly good agreement with the
observed solar differential rotation in both the pole-equator contrast
and also the more conical shaped iso-rotation contours in the mid-latitude
zone.  The more conical iso-rotation contours are achieved by the latitudinal
gradient of entropy imposed at the lower boundary as has been described
in \citet{Rempel:2005, Miesch:etal:2006}.  The latitudinal gradient of entropy
is spread into the bulk of the CZ due to the large thermal diffusivity, and
this latitudinal gradient in the CZ provides the necessary 
balance in the $\phi$ component of the vorticity equation
to allow for a non-cylindrical differential rotation to be established
\citep{Rempel:2005}. Thus the role of the imposed latitudinal entropy
gradient is to change the shape of the iso-rotation contours from cylindrical
to more conical. It is not the reason for the change of differential
rotation from anti-solar (fast rotating pole) to solar-like (faster equator)
between the HD and the magnetic cases, both of which have the same latitudinal
entropy gradient imposed.  That change is brought about by the
change in the direction of Reynolds stress transport of the angular momentum.

\begin{figure}
\centering
\includegraphics[width=0.8\textwidth]{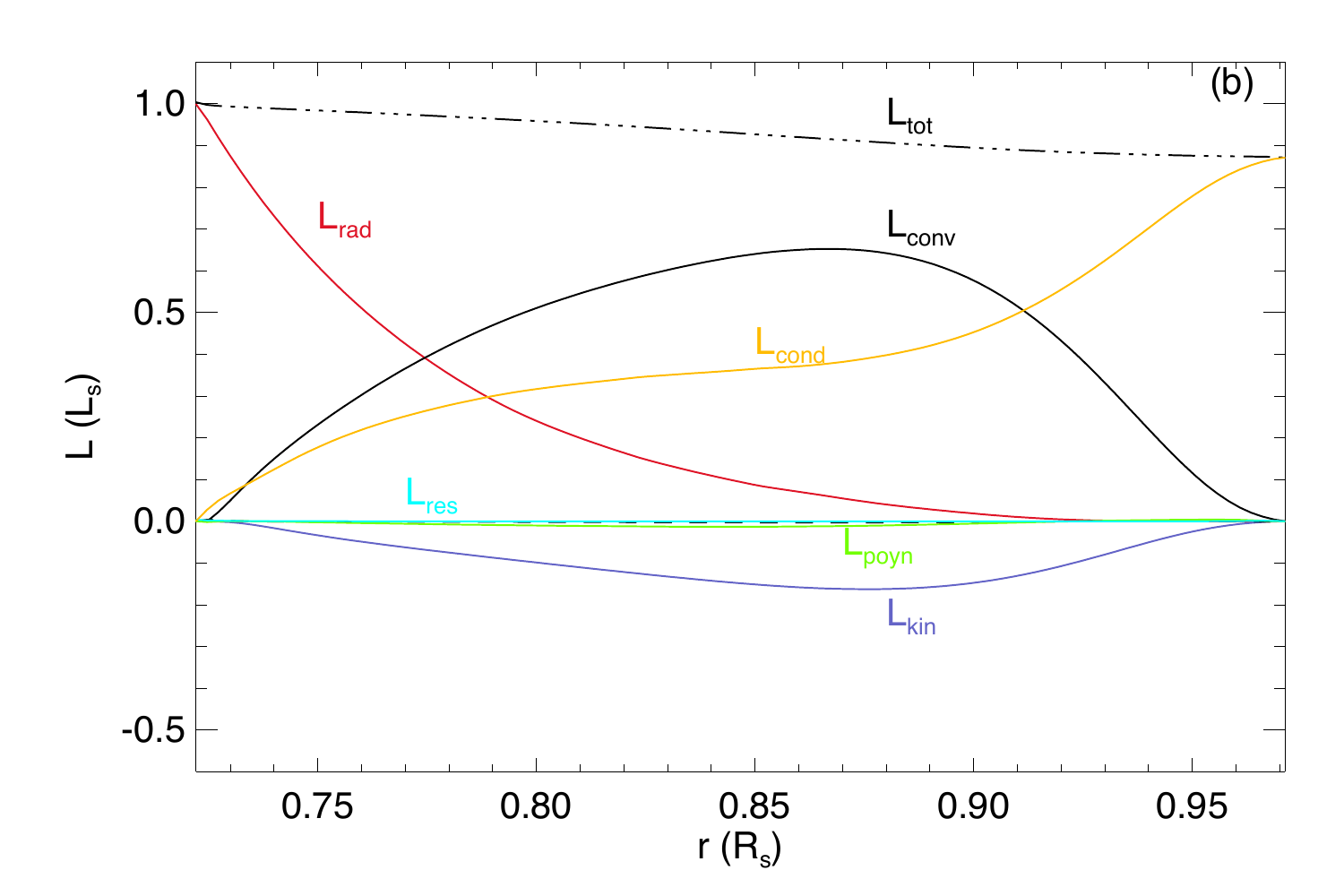}
\caption{The
horizontally integrated energy fluxes due to respectively, radiative diffusion
$L_{\rm rad}$ (red curve), convection $L_{\rm conv}$ (black curve), thermal
conduction $L_{\rm cond}$ (yellow curve), kinetic energy flux 
$L_{\rm kin}$ (blue curve), viscous flux $L_{\rm vis}$ (dashed line),
Poynting flux $L_{\rm poyn}$ (green curve), resistive flux $L_{\rm res}$
(cyan curve), and the sum of all $L_{\rm tot}$ (dash-dotted black curve),
as a function of depth (see text
for the expressions of $L_{\rm rad}$, $L_{\rm conv}$, $L_{\rm cond}$,
$L_{\rm kin}$, $L_{\rm vis}$, $L_{\rm poyn}$, and $L_{\rm res}$).}
\label{fig_heatfluxes}
\end{figure}

\begin{figure}
\centering
\includegraphics[width=0.8\textwidth]{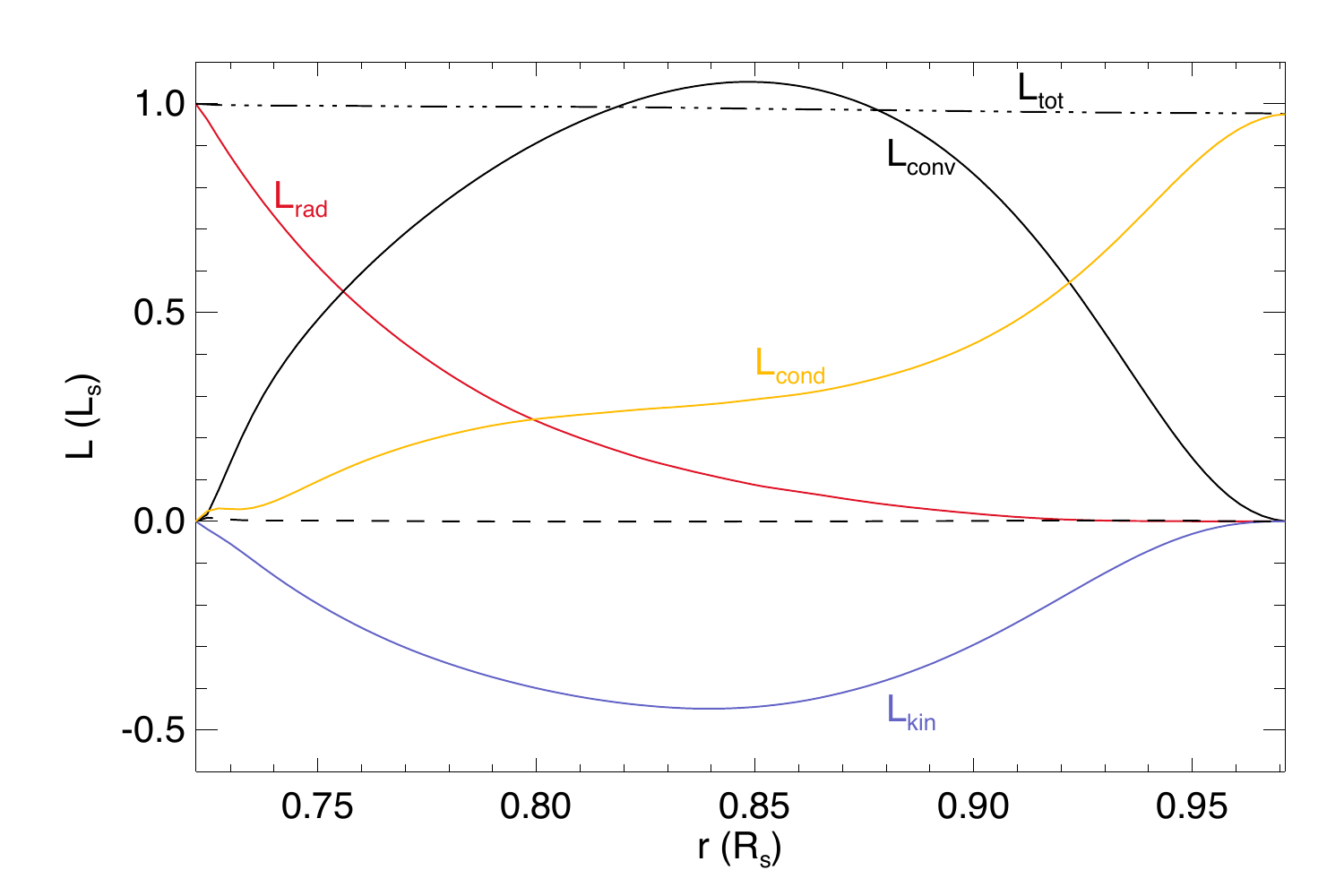} \\
\includegraphics[width=0.8\textwidth]{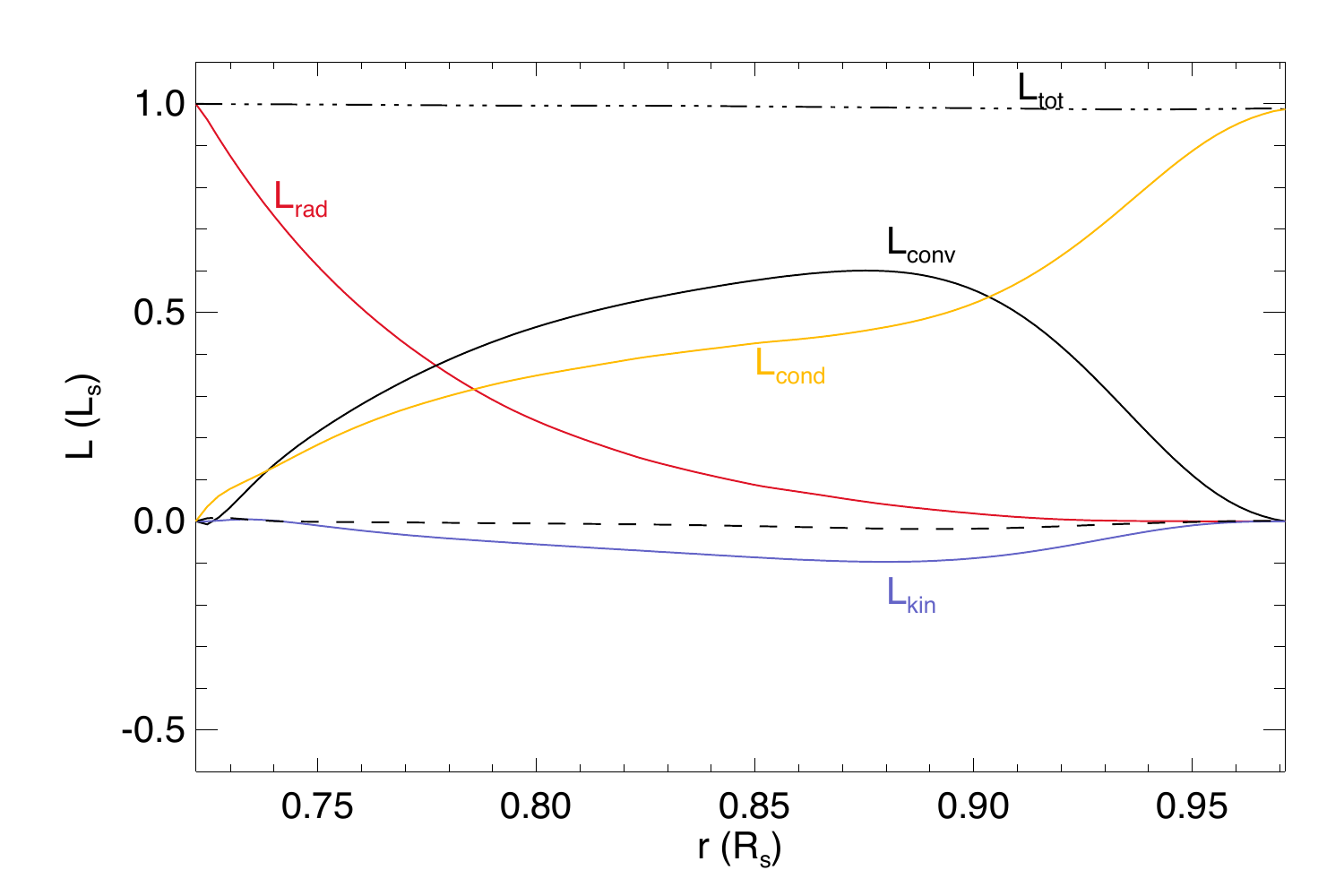}
\caption{Same as Figure \ref{fig_heatfluxes} but for the corresponding
HD case (upper panel) and the HVHD case (lower panel).}
\label{fig_heatfluxes_hdandhdhvis}
\end{figure}

We also note here the possible effect of the small deviation from
total energy conservation in our dynamo simulation as seen in
Figure \ref{fig_heatfluxes}, where the total energy flux $L_{\rm tot}$
exiting the domain at the top is about 13\% less than the input solar
luminosity.  As pointed out in section \ref{subsec:overview}, this
is due to the loss of the magnetic energy dissipated by the numerical
diffusion, which is not put back into the thermal energy and hence does
not have to be carried out by thermal conduction at the top.
This would mean a slightly weakened driving of convection where a reduced
solar luminosity (reduced by up to about $13$\% near the top) is forced
through the domain. The heat fluxes by thermal conduction and convection
would have been slightly higher if the energy conservation were
strictly adhered to.
This however does not affect our finding, that
the magnetic field takes up the role of an enhanced viscosity,
and if it damps the large scale convective motions to the level similar to
that of the HVHD case (with convection carrying roughly 66\% of the solar
luminosity) a solar like differential rotation can be achieved by the
resulting Reynolds stress.
The larger question that remains is how to maintain the solar differential
rotation with a convection that can transport nearly 100\% of the
solar luminosity without resorting to an ad hoc thermal diffusion
(conduction) that carries a substantial fraction (about 36\% at the middle
of the CZ) of the solar luminosity.

\begin{figure}
\centering
\includegraphics[width=0.45\linewidth]{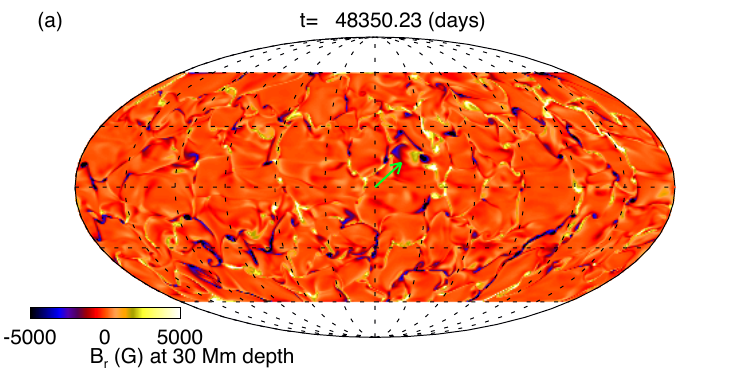}
\includegraphics[width=0.45\linewidth]{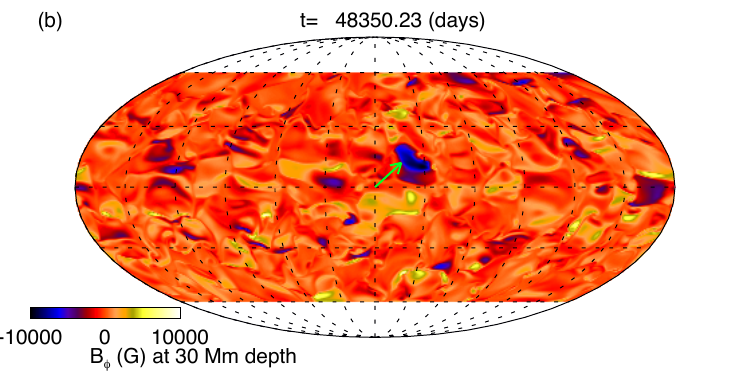} \\
\includegraphics[width=0.45\linewidth]{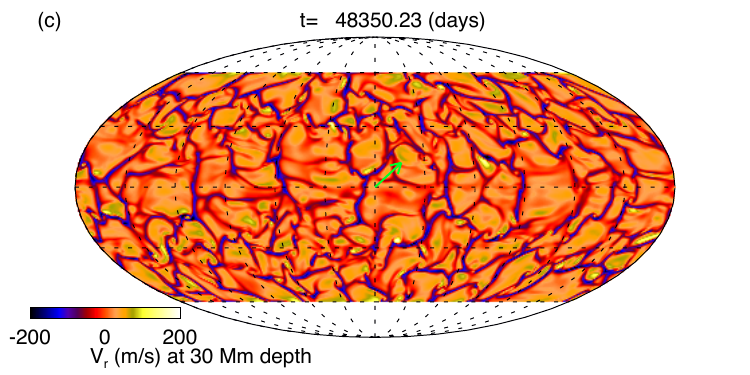}
\includegraphics[width=0.45\linewidth]{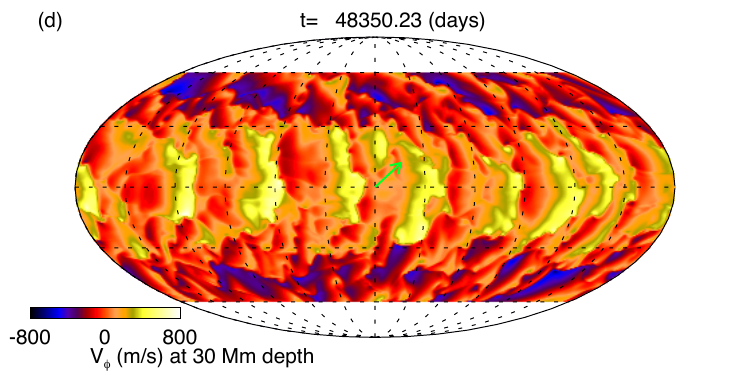} \\
\includegraphics[width=0.45\linewidth]{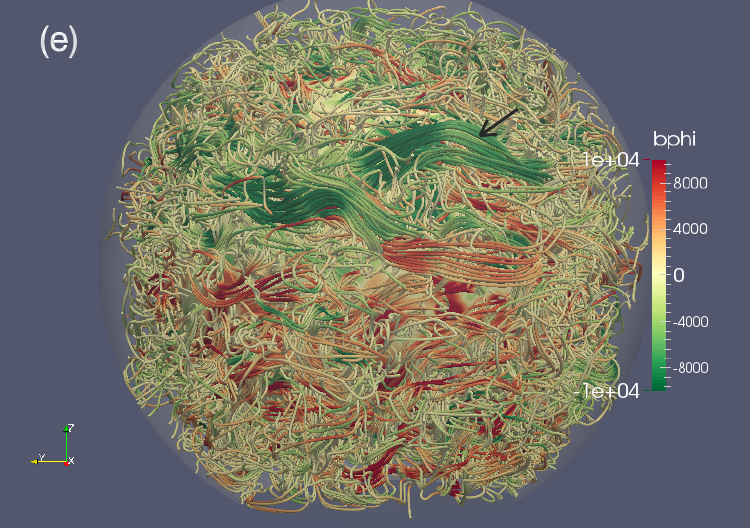}
\includegraphics[width=0.45\linewidth]{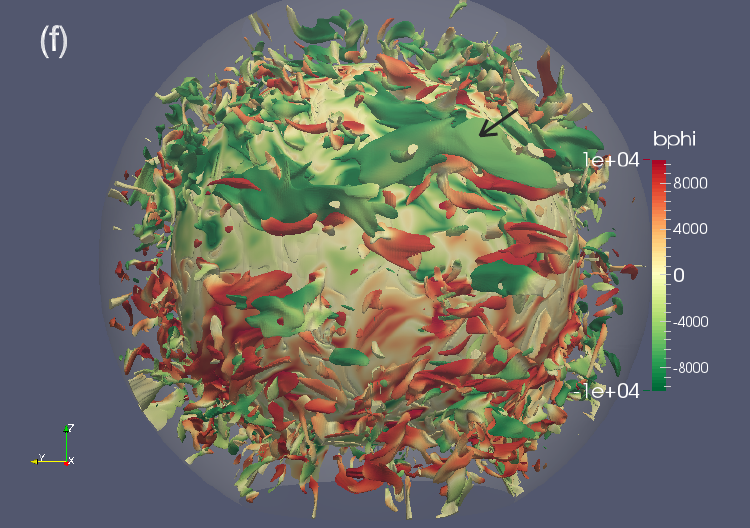}
\caption{Panels (a), (b), (c), and (d) show respectively snapshots of
$B_r$, $B_{\phi}$, $v_r$, and $v_{\phi}$ at a shell slice at the depth
of $30$ Mm below the photosphere, displayed on the full sphere in
Mollweide projection. A movie showing the evolution of $B_r$, $B_{\phi}$,
$v_r$ at the $30$ Mm depth, and also $B_{\phi}$ at a depth near the bottom of
the CZ, over a period of about 13 days centered around the
time instant shown in this Figure is also available in the electronic version.
Panels (e) and (f) show respectively 3D views of the magnetic field lines and
the equipartition field iso-surfaces of $v_a / v_{\rm rms} = 1$
with $v_a$ being the Alfv\'en speed and $v_{\rm rms}$ being the r.m.s.
convective velocity for the corresponding depth.}
\label{fig_emgevent}
\end{figure}

\begin{figure}
\centering
\includegraphics[width=0.45\linewidth]{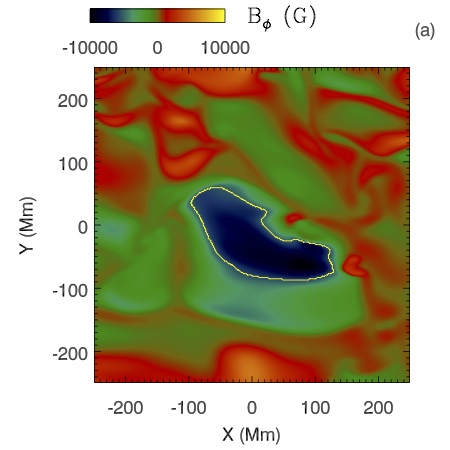}
\includegraphics[width=0.45\linewidth]{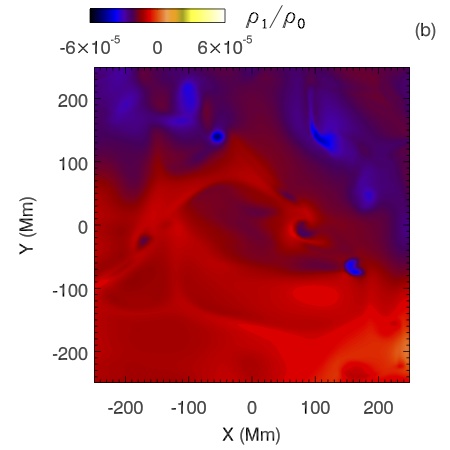} \\
\includegraphics[width=0.45\linewidth]{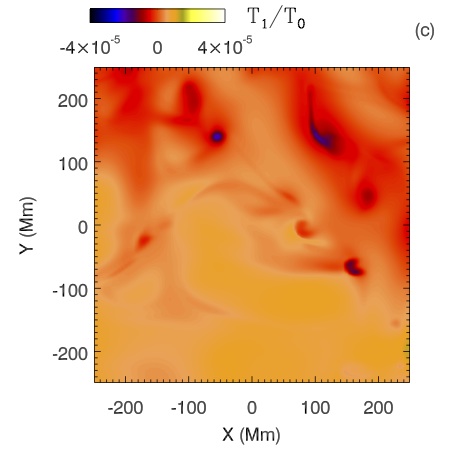}
\includegraphics[width=0.45\linewidth]{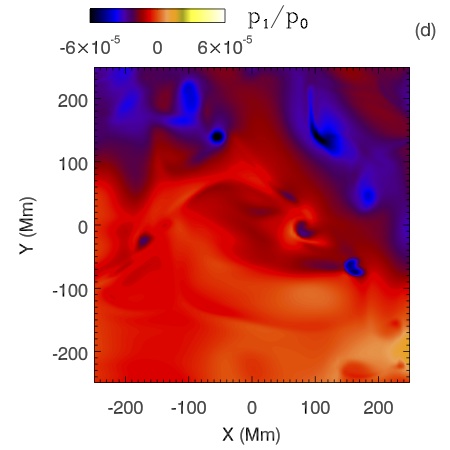}
\caption{Zoomed in view centered on the emerging region on the horizontal surface at the
same depth ($30$ Mm) as shown in Figure \ref{fig_emgevent} with (a) showing $B_{\phi}$ and
a yellow contour marking the emerging region where $B_{\phi} < -5000$ G and the ratio
of the Alfv\'en speed over the r.m.s. convective speed $v_a / v_{\rm rms} > 1$, 
(b) showing density change $\rho_1$ relative to the reference state $\rho_0$ at
that height, (c) showing temperature change $T_1$ relative to the reference state
$T_0$, and (d) showing pressure change $p_1$ relative to the reference
state $p_0$.}
\label{fig_emgzoomview}
\end{figure}

\begin{figure}
\centering
\includegraphics[width=0.40\linewidth]{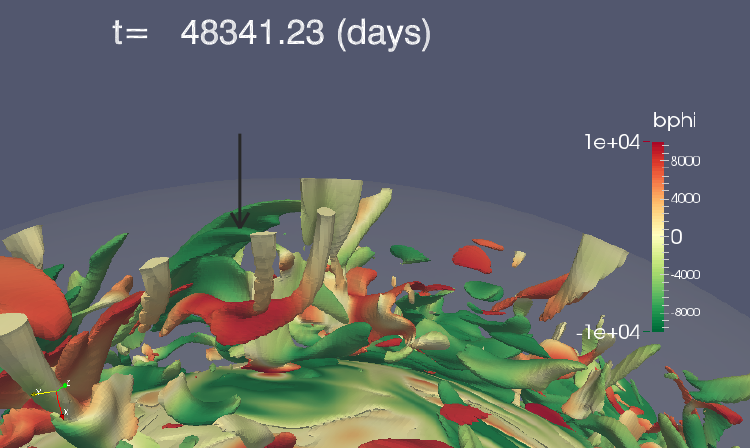}
\includegraphics[width=0.40\linewidth]{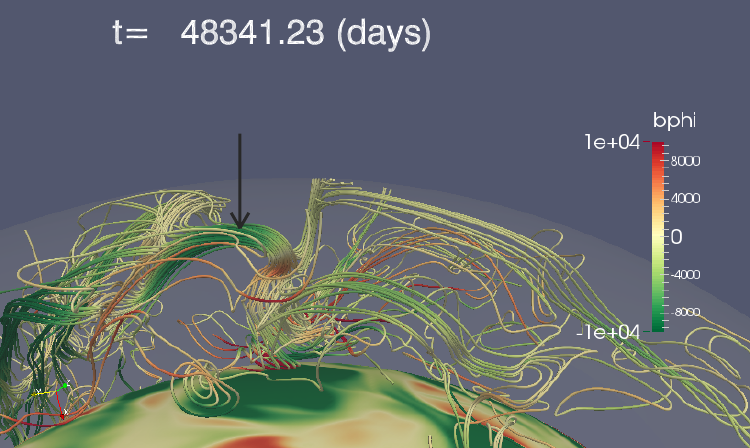} \\
\includegraphics[width=0.40\linewidth]{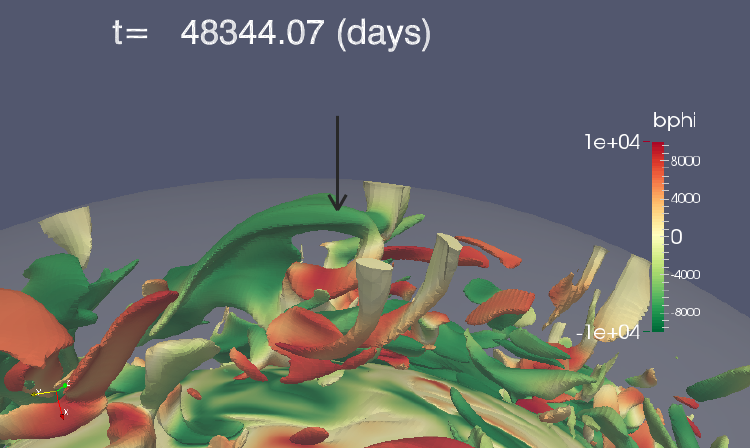}
\includegraphics[width=0.40\linewidth]{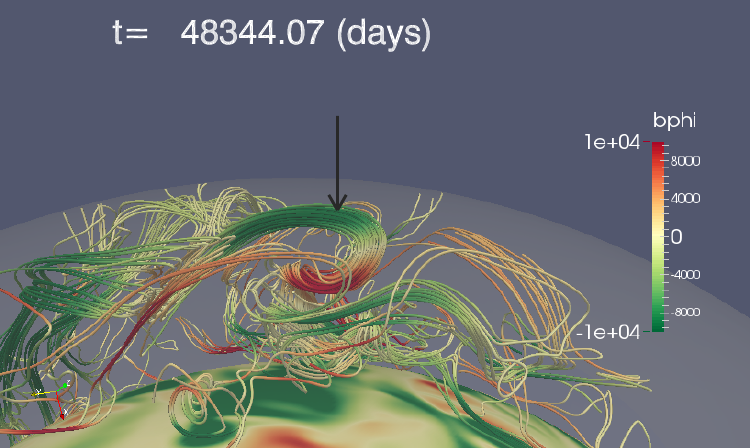} \\
\includegraphics[width=0.40\linewidth]{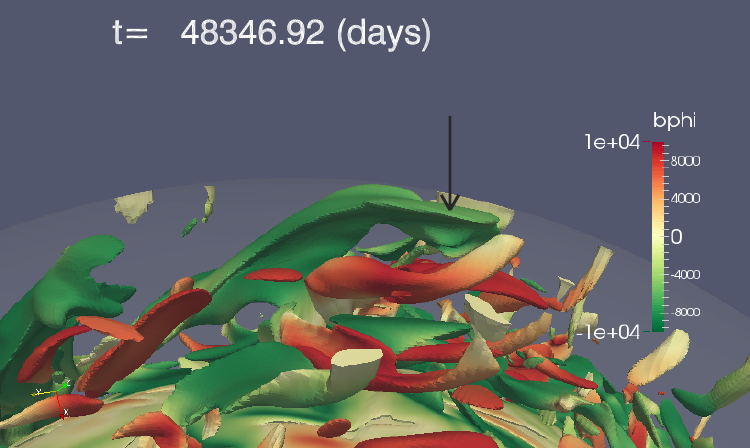}
\includegraphics[width=0.40\linewidth]{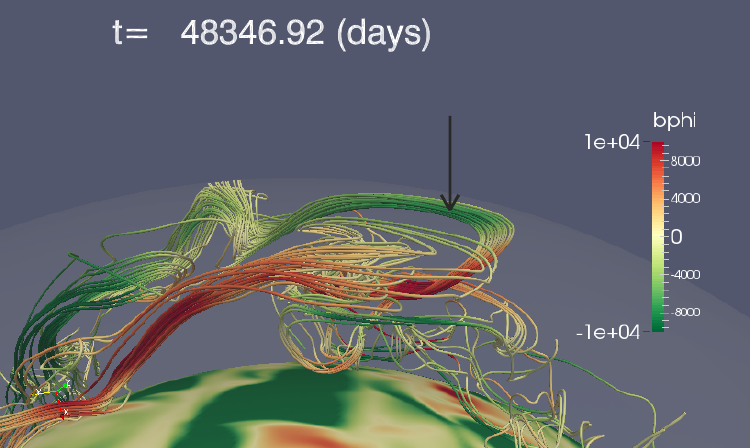} \\
\includegraphics[width=0.40\linewidth]{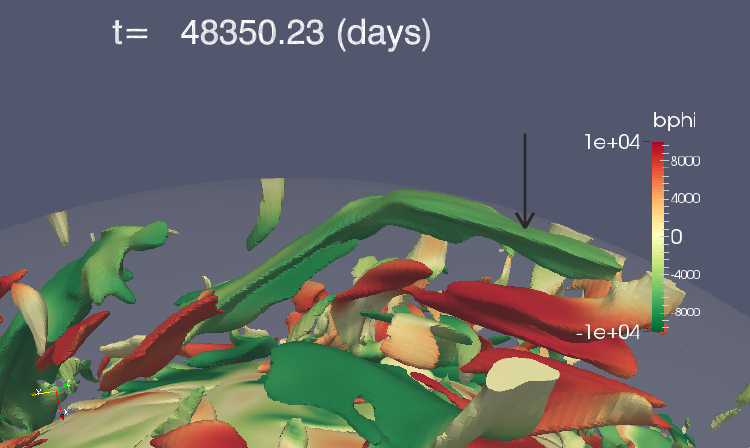}
\includegraphics[width=0.40\linewidth]{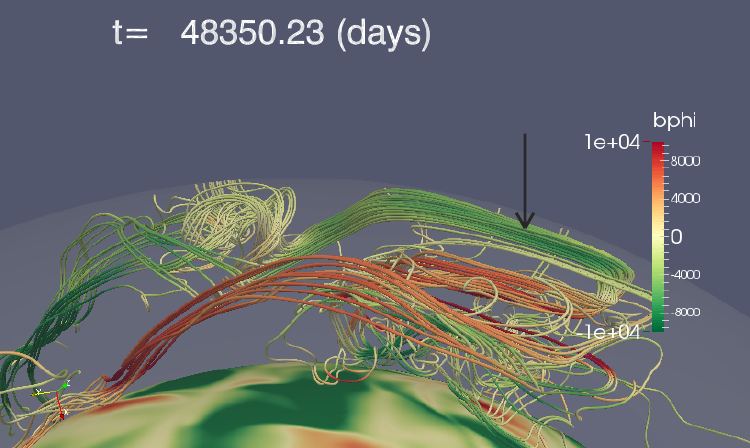} \\
\caption{Left column images show iso-volumes of super-equipartition fields (where
$v_a/v_{\rm rms} > 1$) for a period of 9 days prior to the time of the flux emergence
event shown in Figure \ref{fig_emgevent}, with the arrow marking the evolution of the
emerging flux bundle that produces the flux emergence event shown in 
Figure \ref{fig_emgevent}. Right column images show representative field lines traced
from points in the iso-volume corresponding to the emerging flux bundle.}
\label{fig_shearhairpin}
\end{figure}

In our dynamo simulation the large scale mean toroidal field, antisymmetric
with respect to the equator, is concentrated at the bottom of the CZ, unlike
many of the recent convective dynamo simulations with cylindrical iso-rotation
contours and significantly faster (than solar) rotation rates
\citep[e.g.][]{Nelson:etal:2013,Augustson:etal:2013}, where strong wreath
like toroidal field structures are present in the equatorial region of the
bulk of the CZ. In the 3D magnetic field of the present
simulation (Figure \ref{fig_emgevent}), occasional more
coherent toroidal flux bundles of super-equipartition field strength are
embedded in the turbulent small scale fields, as discussed in section
\ref{subsec:emgflux}.
Some of these super-equipartition flux bundles rise to the surface to
produce active region like flux emergence events. Although these emerging
flux bundles show significant magnetic buoyancy, they are not flux bundles
rising in isolation from the bottom of the CZ, but are product of continued
reconnection and shear amplification by local flows in the CZ. 
There is a preference for the azimuthal field in the strong emerging field
regions to conform to Hale's polarity rule by a ratio of 2.4 to 1 in area,
and a statistical significant mean tilt angle of $7.5^\circ \pm 1.6^\circ$
for the emerging horizontal fields, consistent
with the active region mean tilt.  However, the violation from Hale's rule
is far greater than that of solar active regions. This is because of the very
weak mean field component ($\sim 5$\% of the total magnetic energy) in the
current convective dynamo. It is very likely that our dynamo model is still
significantly over-estimating the giant-cell convective speed
\citep{Hanasoge:etal:2012} and the Sun's dynamo is operating in a
significantly more rotationally constrained regime. This is indicated by the
more rotationally constrained convective dynamo models of
\citet{Kapyla:etal:2012,Augustson:etal:2013},
which are able to achieved a more regular, solar cycle like cyclic behavior and
stronger mean field by
effectively increasing the rotation rate. With a significantly lower magnetic
diffusion (than have been achieved by current global convective dynamo
models), the solar dynamo may be operating in a significantly stronger
field regime with a much more suppressed giant-cell convection, and lower Rossby
number. This may lead to a stronger Reynolds stress transport of angular
momentum that needs to be balanced by a stronger Maxwell's stress and hence
the generation of a stronger mean field.  But the question
remains in regard to how such suppressed convective flows transport the solar
luminosity through the CZ, although the convective energy transport in the
more magnetic buoyancy dominated regime may be quite different.
Furthermore, the inclusion of an overshoot layer below the base of the CZ
may be important for the operation of the solar cycle dynamo. The
convective dynamo model of \citet{Ghizaru:etal:2010,
Racine:etal:2011} has achieved a much stronger large scale mean field
component, with the mean field magnetic energy comparable to that of the
small scale magnetic energy during cycle maxima (compared to the 10\% in
our model), by allowing penetration and storage of the strong mean field 
in the stable overshoot layer.

\acknowledgements
We thank Matthias Rempel and Piyali Chatterjee for discussions and helpful
comments on the paper, and Doug Braun for helpful suggestions on error analysis.
We also thank the anonymous referees for the helpful comments.
This work is supported by the NASA LWSCSW grant NNX13AG04A and NASA HSR
grant NNX10AB81G to NCAR.
NCAR is sponsored by the National Science Foundation.
The numerical simulations were carried out on the Pleiades
supercomputer at the NASA Advanced Supercomputing Division under
project GIDs s1106, s0925.

\clearpage

\clearpage

\clearpage

\clearpage

\clearpage

\clearpage

\clearpage

\clearpage

\clearpage

\clearpage

\clearpage

\clearpage

\clearpage

\end{document}